\documentclass[]{aa}  

\usepackage{amsmath,amssymb}
\usepackage[nopatch]{microtype}
\usepackage{booktabs}
\usepackage{footmisc}

\bibpunct{(}{)}{;}{a}{}{,} 
\usepackage{booktabs} 
\usepackage{threeparttable,longtable}  
\usepackage{makecell}
\usepackage{cprotect}
\usepackage{longtable}
\usepackage{threeparttable}
\usepackage{multicol}
\usepackage{multirow}
\usepackage{textcomp,gensymb}

\usepackage{upgreek}
\usepackage{orcidlink}

\usepackage{rotating}

\usepackage{xpatch}
\usepackage{hyperref}
\hypersetup{
	colorlinks=true,
	breaklinks=true,
	citecolor=blue,
	allcolors=blue,
	frenchlinks=true
}

\makeatletter
\xpatchcmd\NAT@citex
{%
	\@citea\NAT@hyper@{%
		\NAT@nmfmt{\NAT@nm}%
		\hyper@natlinkbreak{\NAT@aysep\NAT@spacechar}{\@citeb\@extra@b@citeb}%
		\NAT@date
	}%
}
{%
	\@citea
	\NAT@nmfmt{\NAT@nm}%
	\NAT@aysep\NAT@spacechar
	\NAT@hyper@{\NAT@date}%
}
{}{}
\xpatchcmd\NAT@citex
{%
	\@citea\NAT@hyper@{%
		\NAT@nmfmt{\NAT@nm}%
		\hyper@natlinkbreak{\NAT@spacechar\NAT@@open\if*#1*\else#1\NAT@spacechar\fi}%
		{\@citeb\@extra@b@citeb}%
		\NAT@date
	}%
}
{
	\@citea
	\NAT@nmfmt{\NAT@nm}%
	\NAT@spacechar\NAT@@open\if*#1*\else#1\NAT@spacechar\fi
	\NAT@hyper@{\NAT@date}%
}
{}{}
\makeatother

\makeatletter
\renewcommand*\aa@pageof{, page \thepage{} of 13}
\makeatother

\begin{document}

\title{A ``MeerKAT-meets-LOFAR'' Study of 2A~0335+096: Discovery of a Complex 500\,kpc Radio Halo in a Disturbed Cool-Core Cluster}

   \author{C.~J.~Riseley\,\orcidlink{0000-0002-3369-1085}\,\inst{1,2},
        R.~J.~van Weeren\,\orcidlink{0000-0002-0587-1660}\,\inst{3},
        L.~Lovisari\,\orcidlink{0000-0002-3754-2415}\,\inst{4},
        R.~Timmerman\,\orcidlink{0000-0001-9404-2612}\,\inst{5,6},
        A.~Botteon\,\orcidlink{0000-0002-9325-1567}\,\inst{7},
        A.~Bonafede\,\orcidlink{0000-0002-5068-4581}\,\inst{8,7},
        A.~Ignesti\,\orcidlink{0000-0003-1581-0092}\,\inst{9},
        N.~Biava\,\orcidlink{0000-0003-1581-0092}\,\inst{7,10},
        G.~Brunetti\,\orcidlink{0000-0003-4195-8613}\,\inst{7},
        B.~Colquhoun\,\inst{11},
        Y.~Perrott\,\orcidlink{0000-0002-6255-8240}\,\inst{11},
        T.~Vernstrom\,\orcidlink{0000-0001-7093-3875}\,\inst{12}
    }
    \authorrunning{C. J. Riseley et al.}
    \titlerunning{MeerKAT-meets-LOFAR: 2A~0335+096}
    \institute{
        Astronomisches Institut der Ruhr-Universit\"{a}t Bochum (AIRUB), Universit\"{a}tsstra{\ss}e 150, 44801 Bochum, Germany\\
        \email{riseley@astro.rub.de}
        \and 
        Ruhr Astroparticle and Plasma Physics Center (RAPP Center), 44780 Bochum, Germany
        \and 
        Leiden Observatory, Leiden University, PO Box 9513, 2300 RA Leiden, The Netherlands
        \and 
        INAF -- IASF Istituto di Astrofisica Spaziale e Fisica Cosmica di
Milano, Via Alfonso Corti 12, 20133 Milano, Italy
        \and 
        Centre for Extragalactic Astronomy,Department of Physics, Durham University, Durham DH1 3LE, UK
        \and 
        Institute for Computational Cosmology, Department of Physics, Durham University, South Road, Durham DH1 3LE, UK
        \and 
        INAF -- Istituto di Radioastronomia, via P. Gobetti 101, 40129 Bologna, Italy
        \and 
        Dipartimento di Fisica e Astronomia, Universit\`a di Bologna, via P. Gobetti 93/2 - IT 40129 Bologna, Italy
        \and 
        Astronomical Institute of the Czech Academy of Sciences, Bo\v{c}n\'{i} II 1401, 14100 Prague, Czech Republic
        \and 
        Thuringer Landessternwarte, Sternwarte 5, 07778 Tautenburg, Germany
        \and 
        School of Chemical and Physical Sciences, Victoria University of Wellington, Wellington 6012, New Zealand
        \and 
        CSIRO Space \& Astronomy, PO Box 1130, Bentley, Western Australia 6102, Australia
    }

      \date{Received: 04~Aug~2026; accepted: 20~Aug~2026; in original form: 03~Jun~2026}

\abstract
    {The emerging picture of radio haloes and mini-haloes in galaxy clusters is more blurred and far less distinct than once thought. While both trace cosmic ray electrons and magnetic fields in the intracluster medium (ICM), our knowledge of the mechanisms powering these sources, their properties and their connection to cluster dynamics, is incomplete. We present new deep MeerKAT L-band (856$-$1712\,MHz) observations of the sloshing cool-core cluster 2A~0335+096, performed as part of the ``MeerKAT-meets-LOFAR'' mini-halo census, as well as reprocessed LOFAR 120$-$168\,MHz LOFAR Two-metre Sky Survey data. We present highly sensitive radio continuum images, from which we perform a spectral analysis and a comparison of the thermal and non-thermal properties using reprocessed archival \textit{XMM-Newton} data. Our MeerKAT data reveal that the known mini-halo is embedded in a large-scale radio halo spanning up to 510\,kpc in this sloshing cluster, although only the inner mini-halo is detected by LOFAR. We find a typical steep spectrum of $\alpha = -1.20 \pm 0.09$ for the mini-halo; the non-detection of the larger-scale halo by LOFAR indicates a spectral index no steeper than $-1.2$. Our results are broadly consistent with an interpretation of turbulent (re-)acceleration powered by sloshing within the cluster potential well. This work emphasises the need for using deep, high-dynamic multi-frequency radio observations in conjunction with high-quality X-ray observations to complete our understanding of the connection between radio halo properties and cluster dynamics.}

\keywords{Galaxies: clusters: individual: 2A0335+096; Radio continuum: general; X-rays: galaxies: clusters}

\maketitle
\nolinenumbers

\section{Introduction}
From the earliest radio-wavelength observations of clusters such as Coma \citep[e.g.][]{Cordey1985_Coma,Brown&Rudnick2011,Bonafede2022_Coma-LOFAR-II} to large cutting-edge samples with the Square Kilometre Array (SKA) Pathfinders and Precursors \citep[e.g.][]{Knowles2022_MGCLS,Botteon2022_LOTSS_Planck,Duchesne2024_EMU-ES-Clusters,Kolokythas2025_MGCLS2} observations of diffuse radio emission in galaxy clusters continue to imply the presence of GeV Cosmic Ray electrons (CRe) and magnetic fields on cluster-volume scales, from cluster-member galaxies to greater than Mpc-scales \citep[e.g.][]{Cuciti2022_megahaloes,Botteon2022_A2255,Rajpurohit2025_RadioHalos}.

Given the short lifetimes of GeV CRe due to synchrotron and inverse Compton (IC) losses, the detection of diffuse radio emission in the intracluster medium (ICM) requires some form of in-situ (re-)acceleration \citep[e.g.][]{Brunetti_Jones_2014}. Classically, these diffuse radio sources can be categorised broadly into two groups: radio relics and radio haloes. 

Radio haloes are canonically Mpc-scale diffuse radio sources that follow the thermal emission from the ICM, show a strong connection between thermal and non-thermal components, and are hosted by around $\sim30\%$ of unrelaxed merging clusters \citep[e.g.][]{Cassano2023_LoTSS_Planck,Cuciti2023_LoTSS-DR2_Planck} although this occurrence rate has a dependence on mass, observing frequency and redshift. It is commonly held that radio haloes are powered by CRe (re-)accelerated to the GeV regime by cluster-scale turbulence following merger events \citep[e.g.][]{Brunetti2001,Petrosian2001,Brunetti_Lazarian_2007}.

Radio mini-haloes fall under the broad umbrella of radio haloes. These are smaller-scale structures typically found in the cores of non-merging (relaxed) galaxy clusters \citep[e.g.][]{Giacintucci2017}. While first observations suggested far smaller sizes for mini-haloes (typically $\lesssim 150$\,kpc) newer observations from the SKA Pathfinders and Precursors have blurred this distinction, with increasingly large ``mini''-haloes being discovered, giant radio haloes reported in relaxed galaxy clusters, and multi-component `mini-halo-plus-halo' structures coexisting in some clusters \citep[see for example][and references therein]{Savini2018,Savini2019,Biava2021_RXCJ1720,Riseley2022_MS1455,Riseley2023_A1413,Bruno2023_A2142,Riseley2024_A2142,vanWeeren2024_Perseus,vanWeeren2026_A1775_A1795}. In light of these blurred lines between radio (mini-)haloes it is time to revisit our taxonomy with a focus on the underlying physical mechanism.

In the case of mini-haloes, the mechanism powering the emission is still debated. Turbulence likely plays a role in the generation of these sources \citep[e.g.][]{Gitti2002,Bravi2016,RichardLaferriere2020} as the presence of mini-haloes correlates with the existence of cold fronts -- discontinuities implying some level of disturbance -- in the ICM \citep[see for example][]{Gitti2002,Biava2024_sloshing_MH}. Recent work by \cite{Giacintucci2024_MH_CF} also demonstrated the need for high-quality X-ray observations outside the bright cores of relaxed clusters, with additional cold fronts confining ``mini''-halo emission reported far from the cluster centre in two cases. This suggests that the entire diffuse emission region in ``mini''-halo systems may be related to large-scale sloshing within the cluster potential well.

An alternative mechanism for generating (mini-)haloes in clusters is the secondary electron (or hadronic) model \citep[e.g.][]{Blasi_Colafrancesco_1999,Pfrommer2004,Keshet_Loeb_2010} whereby CRe are generated through inelastic collisions of Cosmic Ray protons (CRp) with thermal protons in the ICM. CRp would be injected by cluster-member AGN, in particular by the brightest cluster galaxy (BCG) and diffuse to fill the cluster volume due to their longer lifetimes of CRp with respect to CRe. The non-detection of significant $\gamma$-ray emission by \textit{Fermi} \citep[e.g.][]{Ackermann2016_Fermi_Coma,Brunetti2017_Coma,Ignesti2020_MHsample,Adam2021_Fermi_Coma,Osinga2024_A2256,Li_Han_2025_Fermi_Coma} strongly challenges hadronic models for the generation of radio haloes, but observational evidence is lacking for mini-haloes. Furthermore, hybrid models have been proposed \citep[e.g.][]{Brunetti_Lazarian_2011,Zandanel2014} whereby turbulence may act to re-accelerate CRe injected by hadronic collisions.

Broadly, our understanding of the mechanisms powering (mini-)haloes in clusters of galaxies is incomplete. This is largely due to the challenge of generating high dynamic range images of mini-haloes, where analysis of the diffuse emission can easily be hindered by the presence of bright radio sources associated with the BCG. We require statistically-meaningful samples of relaxed clusters with high-quality multi-frequency radio observations in order to study the resolved spectral properties of their diffuse mini-haloes, alongside deep X-ray observations to probe the thermal emission from the ICM, search for signs of disturbance via discontinuities (i.e. cold fronts) that might indicate sloshing, as well as study the connection between the thermal and non-thermal properties in detail.

This paper is the fourth from the ``MeerKAT-meets-LOFAR'' mini-halo census \citep{Riseley2022_MS1455,Riseley2023_A1413,Riseley2024_A2142} which aims to advance our knowledge of the physics of mini-haloes in galaxy clusters. Here we present results on the galaxy cluster 2A~0335+096, which has been historically explored in the radio \citep[e.g.][]{Savini2019,Birzan2020_LOFAR_cavities,Ignesti2021_2A0335}, X-ray \citep[e.g.][]{Sarazin1995_2A0335,Sanders2009_2A0335} and in molecular gas \citep[e.g.][]{Vantyghem2016_2A0335}. In this paper we present new results on the diffuse emission from the ICM using new observations from MeerKAT \citep{Jonas2016} and the LOw-Frequency Array \citep[LOFAR;][]{vanHaarlem2013} in conjunction with re-analysis of archival X-ray data from \textit{XMM-Newton}. We will report on the radio galaxy population associated with this cluster in a future work. Throughout we assume a $\Lambda$CDM cosmology of H$_0 = 73 ~ \rm{km} ~ \rm{s}^{-1} ~ \rm{Mpc}^{-1}$, $\Omega_{\rm{m}} = 0.27$, $\Omega_{\Lambda} = 0.73$. With this cosmology, the angular scale to linear size conversion is 1~arcsec to 1.44~kpc at the cluster redshift \citep[$z = 0.0363$;][]{Sanders2011_clusters_XMM}. We adopt the convention that flux density $S$ is proportional to frequency $\nu$ as $S \propto \nu^{\alpha}$ where $\alpha$ is the spectral index, and we quote all uncertainties at the $1\sigma$ level.

\section{Observations and data processing}

\subsection{Radio data}

\subsubsection{MeerKAT}
2A~0335+096 was observed as part of the ``MeerKAT-meets-LOFAR'' mini-halo census (proposal ID SCI-20210212-CR-01). Two observing runs were performed: the first on 2021-12-16, but due to high winds the run was aborted after around four hours; a second observing run was performed on 2021-12-18. The capture block IDs are 1639674087 and 1639848679, respectively, and the total on-source time was 9.35~hours. Primary calibrators J1939-6342 and J0408-6545 were used to derive bandpass solutions and set the flux density scale; J0521+1638 (3C\,138) was used for polarisation calibration. The secondary calibrator J0323+0534 was used to derive time-varying gains.

Initial calibration was performed in full-polarisation with an in-house recipe using the Common Astronomy Software Applications \citep[\texttt{CASA};][]{McMullin2007_CASA} version \texttt{5.5.0-149.el7} following the same strategy as implemented by \cite{Riseley2025_HCG15}. We refer the reader to that paper for details, but briefly our recipe was as follows: we first removed edge channels and conducted automated sum-threshold flagging using \texttt{tfcrop} and \texttt{rflag}. We set the flux density scale for J1939-6342 (according to the `Stevens-Reynolds 2016' scale) and J0408-6545 using the procedure described on the MeerKAT External Service Desk\footnote{\url{https://skaafrica.atlassian.net/wiki/spaces/ESDKB/pages/1481408634/Flux+and+bandpass+calibration}}. We performed standard continuum calibration, solving for delay and bandpass solutions as well as time-dependent parallel-hand gains. We then performed polarisation calibration using 3C\,138, as described in \cite{Riseley2015_A3667} and \cite{Riseley2025_HCG15}. We then derived flux scale corrections using J1939-6342.

After applying all calibration tables to our visibilities, we imaged 3C\,138 in full polarisation to verify our calibration. The frequency-dependent polarisation properties of 3C\,138 are plotted in Fig.~A1 of \cite {Riseley2025_HCG15} but they are in good agreement with expectations from \cite{Taylor_Legodi_2024}.

Following calibration, we used AOflagger \citep{Offringa2010_AOflagger,Offringa2012_AOflagger} \texttt{v3.0} with a custom strategy to perform RFI excision on our data. Finally, we averaged by a factor 8 in frequency, generating calibrated measurement sets with 465 channels of 1.67~MHz bandwidth before proceeding to self-calibration.

We used \texttt{WSclean} \citep{Offringa2014_WSclean,Offringa2017_WSclean} for imaging and \texttt{killMS} \citep{Tasse2014_killMS,Smirnov2015_killMS} for direction-independent (DI) self-calibration. We performed six rounds of DI self-calibration and imaging, initially solving for phase gains only and subsequently transitioning to amplitude and phase gains, before calibration converged. The wide-field image showed residual direction-dependent (DD) errors that required correction in order to maximise image fidelity.

We performed DD-calibration and imaging using \texttt{facetselfcal}\footnote{\url{https://github.com/rvweeren/lofar_facet_selfcal}} \citep{vanWeeren2021_LOFAR_HETDEX} which employs \texttt{WSclean} for imaging and the Default Preprocessing Pipeline\footnote{\url{https://dp3.readthedocs.io/en/latest/}} (\texttt{DP3}) for self-calibration. We refer the reader to \cite{vanWeeren2021_LOFAR_HETDEX} for detailed discussion and to \cite{Botteon2024_Abell754} for a MeerKAT-specific use case. We first generated an extracted dataset from our DI-calibrated data by subtracting all sources outside a 1.3~degree box around our target, to reduce the computational burden of DD-calibration. We then defined a set of 14 directions in which to derive DD solutions, centred on sources which showed residual DD errors (including one direction centred on our cluster). One round of DD phase-only self-calibration was performed before moving to amplitude and phase mode, as we found that phase-only mode yielded little improvement in the image quality. This suggests that the majority of the DD effects are associated with amplitude errors, consistent with the findings of \cite{vanWeeren2026_A1775_A1795}. We performed 11 rounds of DD self-calibration before our calibration converged, and we show this final wide-field DD-calibrated image in Fig.~\ref{fig:2A0335_fullfield}. We achieve a representative off-source noise of $4.3~\upmu$Jy~beam$^{-1}$ with a resolution of 6.8\,arcsec by 4.7\,arcsec when using \texttt{robust -0.5} weighting.

As a final step, we subtracted all sources outside a 0.45~degree box centred on our target, in order to generate our science images more efficiently. To further refine our calibration, we performed five additional rounds of direction-independent self-calibration before achieving convergence. We show a zoom on this image in the left panel of Fig.~\ref{fig:2A0335_cluster_presub}. We then generated images with different inner \textit{uv}-cuts to isolate sources unrelated to the diffuse emission; these were subtracted from our data using their clean component models. Due to its highly complex structure, we did not attempt to subtract the embedded tailed radio galaxy GB6~B0335+096 \citep[e.g.][]{PatnaikSingh1988,Sebastian2017}.

\subsubsection{LOFAR}
The LOFAR observations used in this paper were performed as part of the LOFAR Two-metre Sky Survey \citep[LoTSS;][]{Shimwell2017_LoTSS_PaperI,Shimwell2019_LOTSS_PaperII,Shimwell2022_LOTSS_PaperIII,Shimwell2026_LOTSS_DR3}. These were performed with the LOFAR High-Band Antennas (HBA) covering frequencies from 120--168\,MHz.

Four LoTSS pointings cover our target: P053+09, P053+11, P055+09 and P056+11. Due to the low Declination of these pointings, each was observed multiple times to achieve the required on-source time of 12 hours for P053+09 and P055+09, and 8 hours for P053+11 and P056+11. We summarise the observing dates for each pointing in Table~\ref{tab:lotss_observations}. We note that while LOFAR data covering 2A~0335+096 have been previously published \citep{Birzan2020_LOFAR_cavities,Ignesti2021_2A0335}, the observations used in this paper are entirely distinct from those datasets.

\begin{table}[htb!]
    \centering
    \caption{LoTSS observations covering 2A~0336+096.}
    \label{tab:lotss_observations}
    \renewcommand{\arraystretch}{1.15}
    \footnotesize
    \begin{tabular}{c | c c | c}
        Pointing & RA       & Dec.      & Observing dates \\
                 & [deg] & [deg]  &       \\
        \hline\hline
        \multirow{2}{*}{P053+09}	&   \multirow{2}{*}{53.3179}	&   \multirow{2}{*}{8.786}	    &   2024-01-18, 22;  \\
            &   &   &   2024-02-08, 22, 23, 24\\
        \hline
        \multirow{2}{*}{P055+09} &   \multirow{2}{*}{55.9372}	&   \multirow{2}{*}{8.807}   &   2024-02-09, 10, 25, 26; \\
            &   &   &   2024-03-11; 2024-07-27 \\
        \hline
        P053+11 &   53.3983	&   11.297  &   2018-01-04; 2023-11-23, 27  \\
        \hline
        \multirow{2}{*}{P056+11} &   \multirow{2}{*}{56.0380}	&   \multirow{2}{*}{11.318}  &   2023-10-10; 2023-11-20, 21, \\
            &   &   &   2023-11-29, 30; 2023-12-07
    \end{tabular}
\end{table}

The LoTSS data were processed using the \textsc{ddf-pipeline}\footnote{\url{https://github.com/mhardcastle/ddf-pipeline}} \citep[see e.g.][and references therein]{Shimwell2022_LOTSS_PaperIII} as part of the LOFAR Surveys Key Science Project. Following wide-field DI and DD calibration, a region around the target was extracted and further self-calibrated using the techniques described in \cite{vanWeeren2021_LOFAR_HETDEX}. 

We started with DI self-calibration on the extracted region, solving for both \texttt{tecandphase} on short timescales and amplitude and phase on longer timescales, similar to the approach used by \cite{Botteon2022_LOTSS_Planck}. A total of nine rounds of self-calibration were performed during the extraction process before convergence was reached. The resulting image, however, still showed clear DD errors across the extracted region. 

We therefore proceeded with an additional DD calibration step using \texttt{facetselfcal} with seven directions. This follows the general approach described in \cite{vanWeeren2026_A1775_A1795}. The calibration consisted of scalar phase-only solutions on timescales of 1--2~min and combined scalar amplitude and phase solutions on timescales of 1~hr. Again, nine rounds of self-calibration were performed, which successfully removed the remaining DD errors.

\begin{figure*}
    \centering
    \includegraphics[width=16cm]{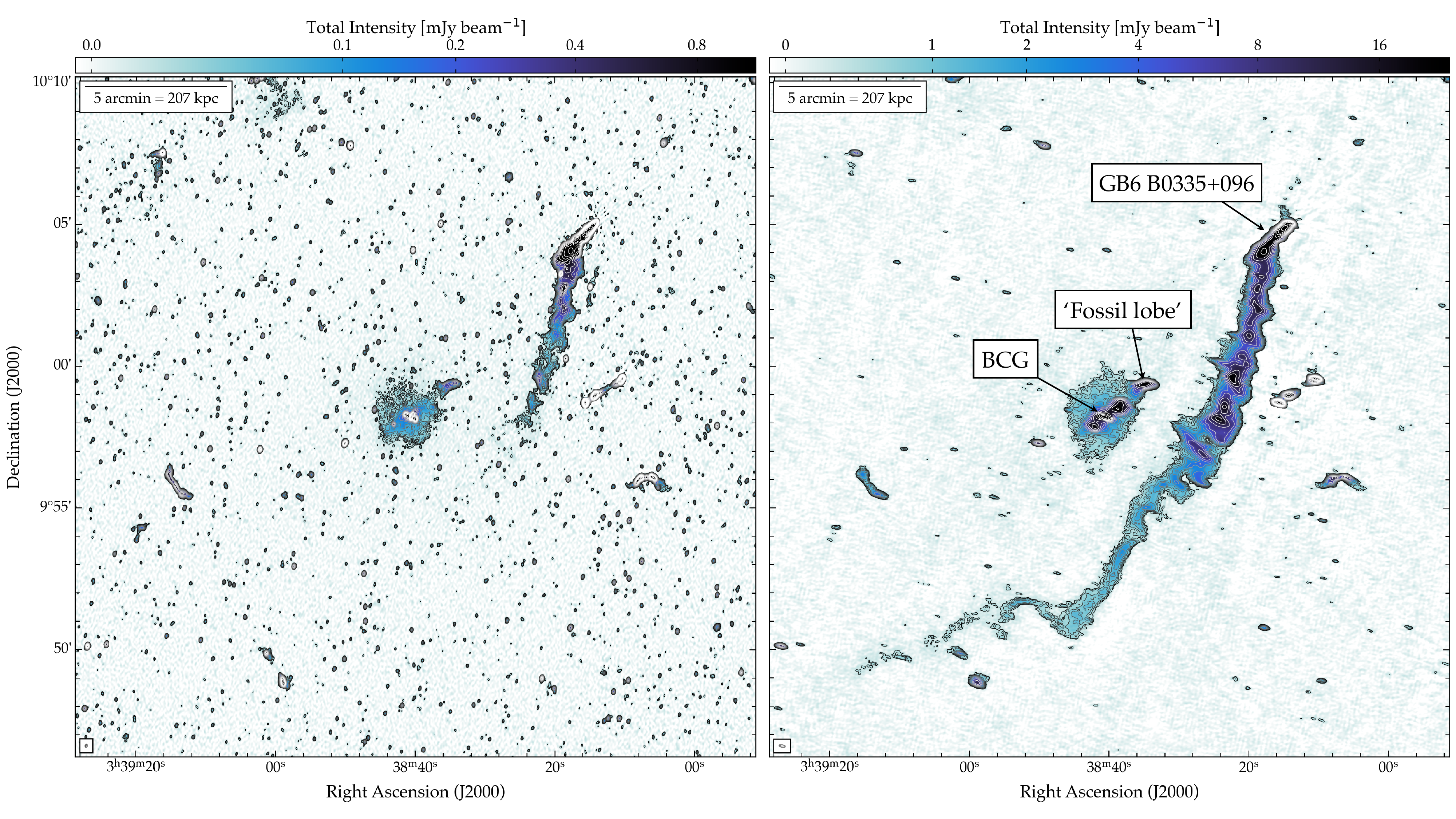}
    \caption{Radio continuum images of 2A~0335+096 as seen by MeerKAT at 1286\,MHz (left) and LOFAR at 145\,MHz (right). The off-source RMS noise is $\sigma = 4.5 ~ \upmu$Jy~beam$^{-1}$ at 1283\,MHz and $\sigma = 95 ~ \upmu$Jy~beam$^{-1}$ at 143\,MHz, where the restoring beam is shown in the lower-left corner of each subplot. The restoring beams are 6.8\,arcsec by 4.7\,arcsec at a PA of 170\,deg for the MeerKAT image and 13.3\,arcsec by 5.2\,arcsec at a PA of 79\,deg for the LOFAR image. Contours start at $4\sigma$ and scale by a factor $\sqrt{2}$. For reference, GB6~B0335+096 is the extended tailed radio galaxy that runs from the north-west of the image to the south/south-east.}
    \label{fig:2A0335_cluster_presub}
\end{figure*}

We then generated images at 6~arcsec resolution and followed established routines to evaluate the scaling factor required to align our data with the LoTSS flux density scale \citep[see for example][]{Hardcastle2016,Shimwell2019_LOTSS_PaperII}. For our data, we found that an average scaling factor of 1.47 was required to align our extracted images with the LoTSS flux density scale. The LOFAR image is shown in the right-hand panel of Fig.~\ref{fig:2A0335_cluster_presub}. Finally, as with our MeerKAT observations, we generated images with different inner \textit{uv}-cuts to isolate and subtract sources unrelated to the diffuse emission associated with the ICM of 2A~0335. We similarly did not subtract the tailed radio galaxy GB6~B0335+096.

\subsection{X-ray data: XMM-Newton}
Our reduction scheme is consistent with that adopted in \cite{Riseley2025_HCG15}. Here we provide a short description of the main steps.
The Observation Data Files (ODFs) were retrieved from the XMM–Newton archive and reprocessed with the XMM–Newton Science Analysis System (SAS) v19.1.0, using the latest calibration files available at the time of the analysis. 
The calibrated event files were produced with the standard tasks {\it emchain} and {\it epchain}. We selected single-quadruple pixel events for MOS (i.e., PATTERN$<$13) and single-double pixel events for pn (PATTERN$<$5), and also performed bright pixels and hot columns removal
(i.e., FLAG==0). 
Periods affected by soft-proton flares were filtered using the ESAS tools {\it mos-filter} and {\it pn-filter}. 
CCDs in anomalous state were excluded following \citep{kun08}. 
Out-of-time (OoT) events were generated and subtracted from pn images and spectra after appropriate rescaling.
Point-like sources were detected using
the task {\it edetect$\_$chain} and excluded from the event files.
Images were extracted in the 0.7-2.0 keV band, background-subtracted and vignetting-corrected. 
For display purposes, point-source regions were refilled using {\it dmfilth} in CIAO.
Spectral fitting was performed in XSPEC (\citealt{Arnaud1996}, v12.11) in the 0.5–12 keV (MOS) and 0.5–14 keV (pn) bands. The emission was modelled with an absorbed APEC thermal plasma model assuming the solar abundances of \cite{Asplund2009}.
The absorbing column density was fixed to $2.62\times10^{20} ~ {\rm cm^{-2}}$.
The best-fit parameters were obtained by minimizing the C-statistics.
The background modelling follows the approach described in \cite{LL19}, with the updates highlighted in \cite{Lovisari2024_CHEX-MATE}.

\section{Results}
\begin{figure*}
    \begin{center}
    \includegraphics[width=0.85\textwidth]{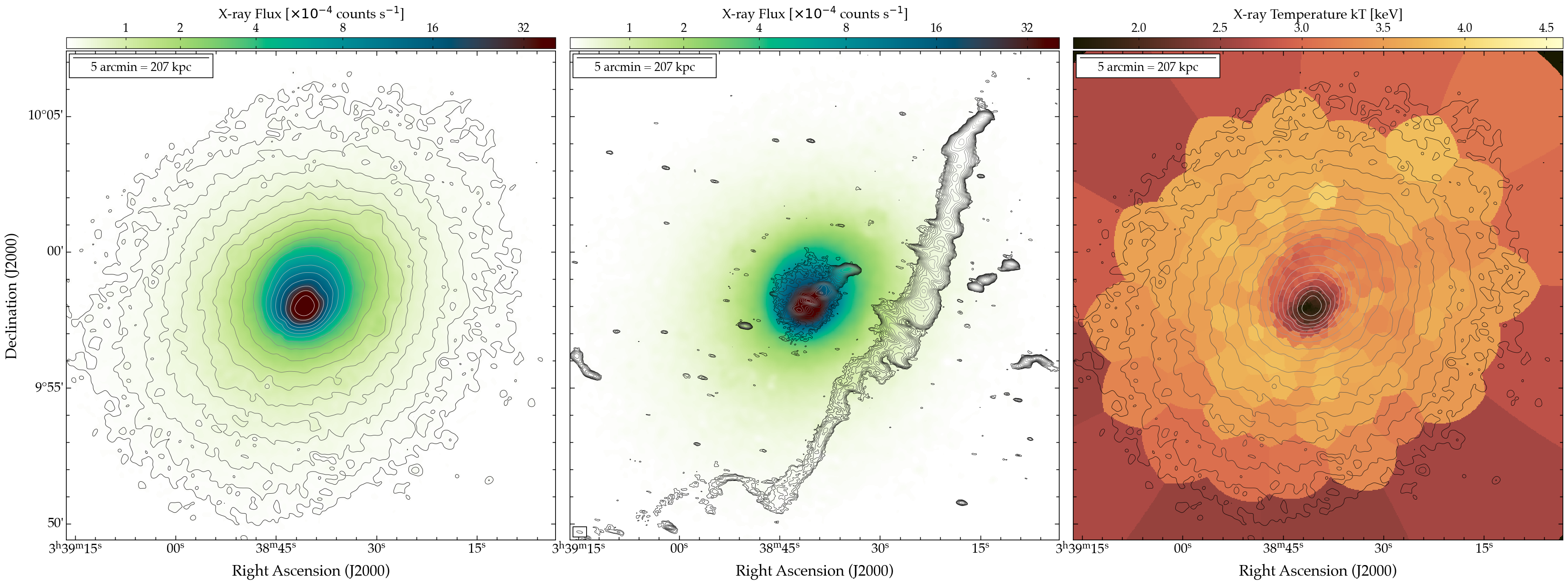}
    \caption{X-ray maps of 2A~0335+096 as produced from our \textit{XMM-Newton} data after smoothing with a Gaussian of 7 pixels FWHM. Contours show the X-ray surface brightness in the left-hand panel and LOFAR surface brightness (as per Fig.~\ref{fig:2A0335_cluster_presub}) in the central panel. X-ray surface brightness contours start at the $3\sigma$ level, where $\sigma = 4.1 \times 10^{-6}$ counts s$^{-1}$ for our \textit{XMM-Newton} image. The right-hand panel shows the X-ray temperature map derived from our \textit{XMM-Newton} data, overlaid with surface brightness contours as per the top-left panel.}
    \label{fig:2A0335_cluster_presub_xray}
    \vspace{-3mm}
    \end{center}
\end{figure*}

Fig.~\ref{fig:2A0335_cluster_presub} presents our radio wavelength images of 2A~0335+096 as seen by MeerKAT at 1283\,MHz (left) and LOFAR at 145\,MHz (right) produced using \texttt{robust -0.5} weighting and prior to subtracting radio sources that are superimposed on the diffuse radio emission. All radio images are generated using a common inner \textit{uv}-cut of $80\lambda$ to ensure consistent scale size sensitivity, corresponding to a maximum recoverable angular scale of around 43\,arcmin. Our \textit{XMM-Newton} image is shown in Fig.~\ref{fig:2A0335_cluster_presub_xray}, along with the temperature map derived from these data.

Many of the radio sources identified in previous works are visible in Fig.~\ref{fig:2A0335_cluster_presub}, including the diffuse mini-halo surrounding the BCG and its lobes, the `fossil/relic lobe' (as referred to by \citealt{Giacintucci2019} and \citealt{Ignesti2021_2A0335} respectively) and the giant head-tail radio galaxy GB6~B0335+096 \citep[e.g.][]{PatnaikSingh1988,Sarazin1995_2A0335,Sebastian2017,Ignesti2021_2A0335}. As visible in the MeerKAT image in particular, the density of radio sources in the region around the mini-halo is high, requiring the careful isolation and subtraction mentioned earlier in order to remove their contribution from the diffuse emission of the mini-halo.

\begin{figure*}
    \sidecaption
    \includegraphics[width=12cm]{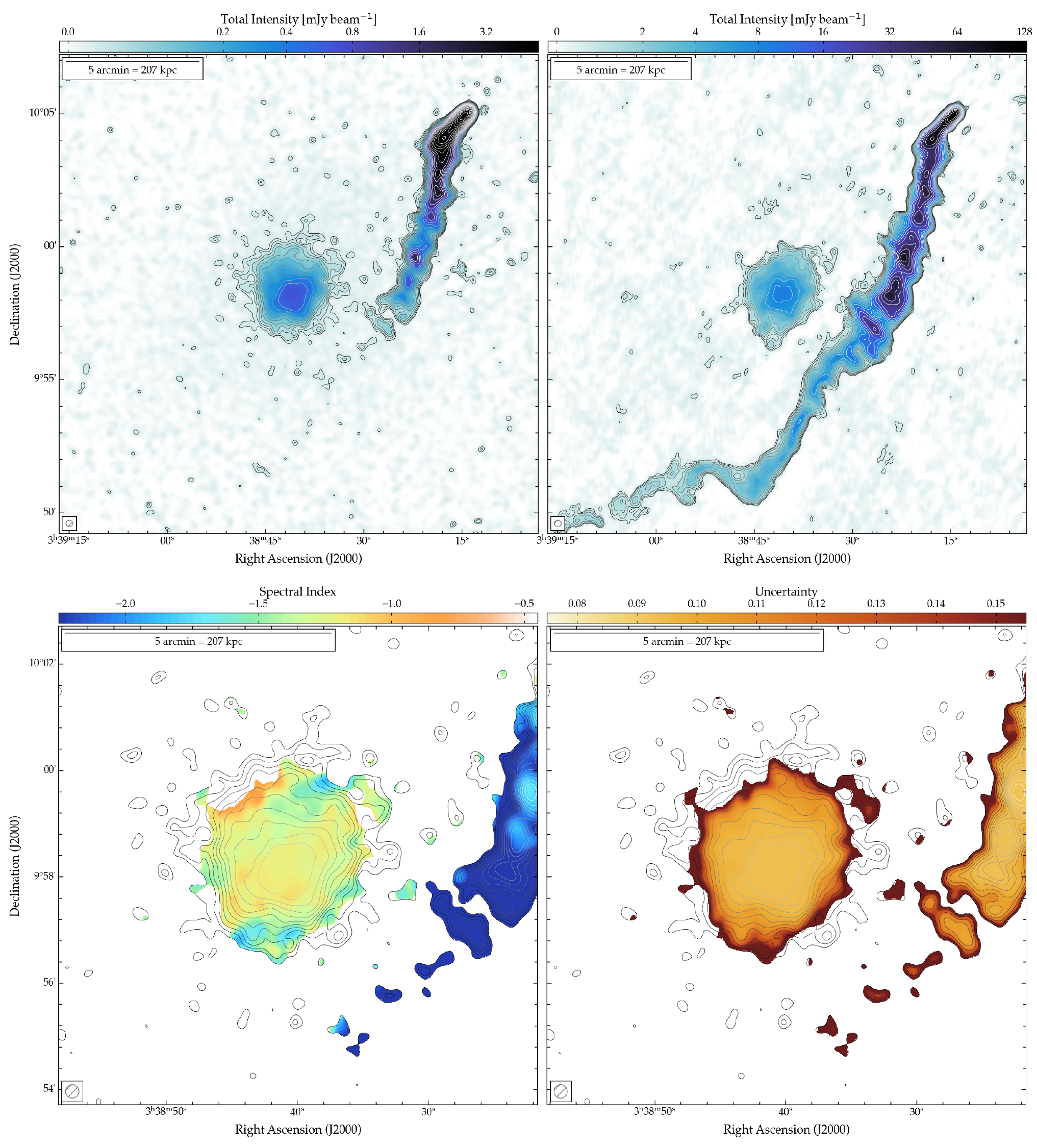}
    \caption{Source subtracted images of 2A~0335+096 at 15\,arcsec resolution. The upper panels show our MeerKAT image at 1283\,MHz and LOFAR image at 145\,MHz. The lower panels show the spectral index and associated uncertainty (zoomed to show the diffuse mini-halo) overlaid with MeerKAT contours. Contours start at $3\sigma$ and scale by $\sqrt{2}$, where the rms noise is $\sigma = 7.3 ~ (175) ~ \upmu$Jy~beam$^{-1}$ for the MeerKAT (LOFAR) images.}
    \label{fig:2A0335_diffuse_15asec}
\end{figure*}

\begin{figure*}
    \sidecaption
    \includegraphics[width=12cm]{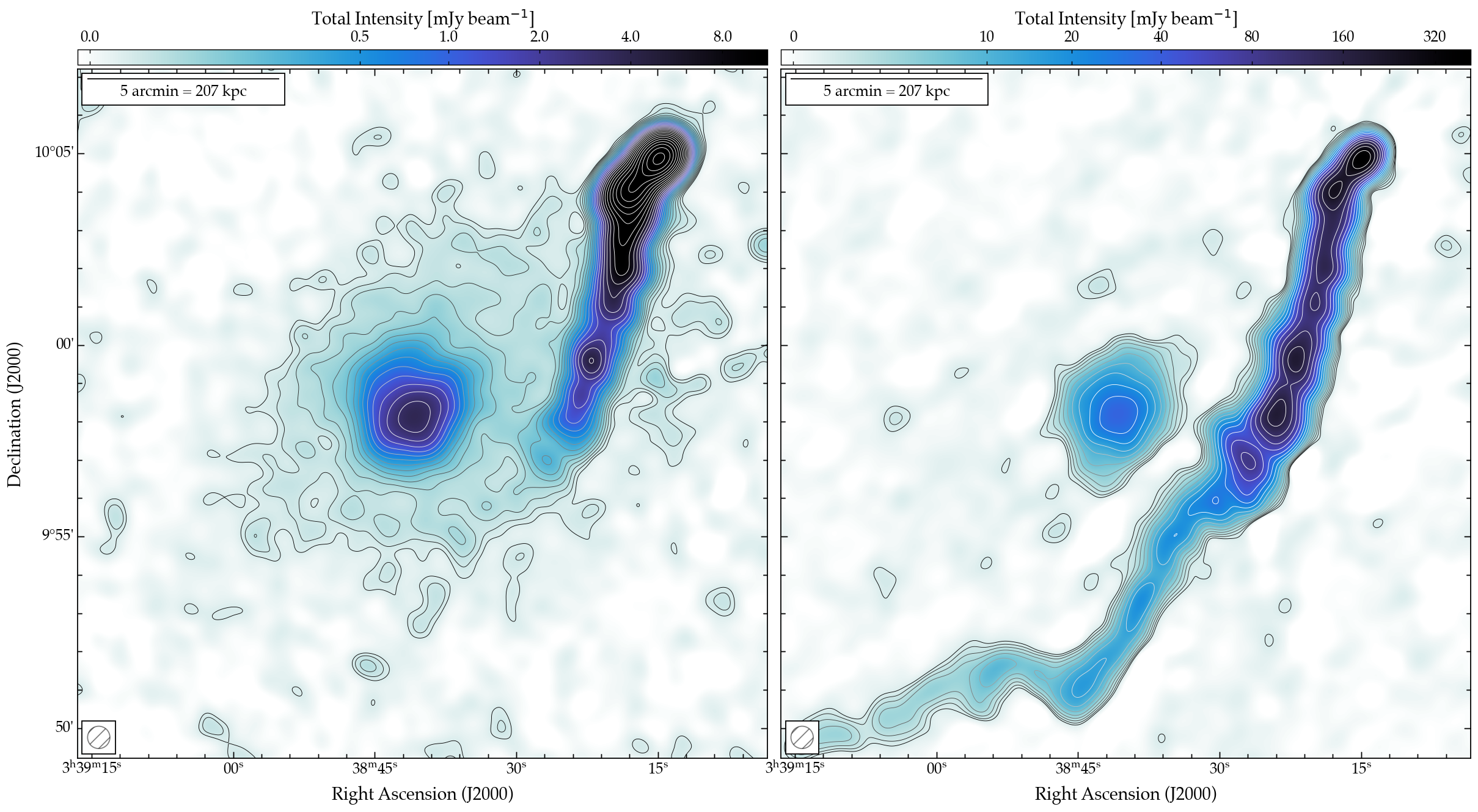}
    \caption{Source subtracted images of 2A~0335+096 at 35\,arcsec resolution. Panel (a) shows the MeerKAT image at 1283\,MHz; panel (b) shows the LOFAR image at 145\,MHz. Contours start at $3\sigma$ and scale by $\sqrt{2}$, where the rms noise is $\sigma = 11 ~ (419) ~ \upmu$Jy~beam$^{-1}$ for the MeerKAT (LOFAR) images.}
    \label{fig:2A0335_diffuse_35asec}
\end{figure*}

In this paper we focus on the diffuse emission associated with the ICM of 2A~0335+096. We will discuss the results on the radio galaxies associated with 2A~0335+096 in another work. We created two sets of lower-resolution source-subtracted images from our MeerKAT and LOFAR data with resolutions of 15\,arcsec and 35\,arcsec, respectively corresponding to a physical resolution of around 10\,kpc and 24\,kpc. We present our 15\,arcsec resolution images as well as the corresponding spectral index map in Fig.~\ref{fig:2A0335_diffuse_15asec}; the 35\,arcsec resolution images are shown in Fig.~\ref{fig:2A0335_diffuse_35asec}.

\subsection{Images at 15 arcsec resolution}
From Fig.~\ref{fig:2A0335_diffuse_15asec} we see that the central mini-halo is well-recovered by both MeerKAT and LOFAR at 15\,arcsec resolution. The diffuse emission is asymmetric, with a greater extension towards the north than the south. Overall the emission spans an area around 4.3\,arcminutes in diameter (corresponding to a linear size around 179\,kpc) and shows a similar extent at both 1283\,MHz and 145\,MHz. This is somewhat larger than previously reported at similar frequencies \citep[$\sim70$\,kpc;][]{Giacintucci2019,Ignesti2021_2A0335} likely owing to the greater sensitivity of our data. However, the general morphology of the diffuse emission at 15\,arcsec resolution is similar to that reported previously.

Integrating over the extent of the mini-halo, we derive an integrated flux density of $S_{145} = 373.7 \pm 37.5$~mJy from our LOFAR image and $S_{1283} = 25.6 \pm 1.3$~mJy from our MeerKAT image. This yields an integrated spectral index of $\alpha_{\rm int} = -1.23 \pm 0.05$ between 1283\,MHz and 145\,MHz, consistent with the spectral index estimation of $\alpha = -1.2 \pm 0.1$ by \cite{Ignesti2021_2A0335}.

In the lower panels of Fig.~\ref{fig:2A0335_diffuse_15asec} we see the spectral index map of the mini-halo at 15\,arcsec resolution. The spectral index is relatively steep with a median value of $\langle \alpha \rangle = -1.20 \pm 0.09$, and is relatively uniform across much of the extent of the mini-halo. As with some (mini-)haloes, we do not see clear evidence of radial steepening \citep[e.g.][]{Riseley2022_MS1455,vanWeeren2026_A1775_A1795} although this is in contrast to others which show evidence of radial steepening \citep[e.g.][]{Biava2021_RXCJ1720,Riseley2023_A1413,Giacintucci2024_MH_CF,Hoang2025_RXJ1347,Trehaeven2025_A3558}. 

While there is no clear global trend, we note some substructure in the spectral index distribution. Toward the north-east of the mini-halo we identify a tentative flattening, where we see a contiguous region of spectral index $\langle \alpha \rangle = -0.91 \pm 0.14$. This tentative flattening may indicate particle acceleration due to a propagating AGN shock, as reported for RBS~797 \citep{Bonafede2023_RBS797,Ubertosi2023_RBS797}. Conversely, toward the north-eastern and southern edges of the mini-halo we see regions with steeper spectra, namely $\langle \alpha \rangle = -1.65 \pm 0.13$ and $\langle \alpha \rangle = -1.56 \pm 0.14$. This steepening may reflect increased turbulence injected by the passage of the head-tail radio galaxy GB6~B0335+096 through the ICM. Further study would be required to investigate these scenarios, which we will return to in our forthcoming analysis of the radio galaxy population of 2A~0335+096.

\subsection{Images at 35 arcsec resolution}
From our 35\,arcsec resolution images in Fig.~\ref{fig:2A0335_diffuse_35asec} however, we see a strikingly different picture. While the extent of the emission recovered by LOFAR is similar to that seen at 15\,arcsec resolution, MeerKAT recovers diffuse radio emission spanning a much larger volume. The diffuse emission spans around 12.3\,arcminutes by 8.9\,arcminutes, corresponding to a physical size of around 510\,kpc by 370\,kpc. The emission is also asymmetric, filling the volume between the cluster centre and the tailed radio galaxy GB6~B0335+096 to the north-west (around 6.8\,arcmin, or 282\,kpc) but extending a lesser distance to the south-east (5.6\,arcmin or 232\,kpc). Perpendicular to this axis, the emission is more symmetric, spanning 3.7\,arcmin (153\,kpc) to the north-east and 4.0\,arcmin (164\,kpc) to the south-west, although we note that this south-western extent includes the emission co-located with the ultra-steep-spectrum tail from GB6~B0335+096 detected by LOFAR. While MeerKAT does not clearly recover emission associated with this component of the tail, we cannot rule out contamination in this sector.

We derive two measures of the integrated flux density for the extended ``mini''-halo by integrating the emission above the contiguous $3\sigma$ contour in Fig.~\ref{fig:2A0335_diffuse_35asec}, excluding contribution from the extended tail of GB6~B0335+096. The more conservative measure also excludes emission seen in the region where we know the tail to extend in our LOFAR image, and here we derive an integrated flux density of $S_{\rm 1283\, MHz} = 34.8 \pm 1.8$~mJy. The less conservative measure assumes that this emission provides a negligible contribution to the diffuse halo, and includes the region immediately beyond the extent of the tail seen in Fig.~\ref{fig:2A0335_diffuse_15asec}; here we derive an integrated flux density of $S_{\rm 1283\, MHz} = 36.1 \pm 1.8$~mJy. These two measures are consistent, suggesting that any residual steep-spectrum emission associated with the tail of GB6~B0335+096 in this region does not provide a significant contribution to the total flux density.

The non-detection of the larger-scale emission by LOFAR does not necessarily imply a flat spectrum. From our MeerKAT image at 1283\,MHz we measure a median surface brightness of around $60 ~ \upmu$Jy beam$^{-1}$ for the outer component; compared to a $2\sigma$ limit from our LOFAR image this would suggest a lower-limit of $\alpha \gtrsim -1.2$, consistent with the spectral index observed for the central region of the ``mini''-halo. As such, we can only rule out an ultra-steep-spectrum for the larger-scale emission, and much deeper LOFAR observations would be required in order to detect this emission at 145\,MHz. Overall, the extent of the ``mini''-halo recovered by MeerKAT at 35\,arcsec resolution is consistent with the emerging picture of extended diffuse emission in sloshing relaxed clusters \citep[e.g.][]{Savini2018,Savini2019,Biava2021_RXCJ1720,Riseley2022_MS1455,Giacintucci2024_MH_CF,Trehaeven2025_A3558,Hoang2025_RXJ1347,Kolokythas2025_MGCLS2,vanWeeren2026_A1775_A1795}.

\section{Analysis}

\subsection{Point-to-point correlations}
It is well-established that in many radio (mini-)haloes, the non-thermal synchrotron emission (traced by radio observations) and the thermal Bremsstrahlung emission (traced by X-ray observations) are well-correlated. Following established techniques \citep[e.g.][]{Govoni2001,Rajpurohit2023_A2256-Halo,Riseley2022_A3266,vanWeeren2026_A1775_A1795} we studied the point-to-point correlations between radio and X-ray surface brightness, as well as between both radio and X-ray surface brightness and X-ray temperature. The latter of these correlations is less well-explored as it also relies on the availability of (very) high signal-to-noise X-ray data from which reliable X-ray temperatures can be derived. 

We extracted surface brightness and temperature measurements across the cluster volume from our images at 35\,arcsec resolution. We smoothed our \textit{XMM-Newton} image with a Gaussian kernel of 35\,arcsec FWHM for the comparison with our 35\,arcsec resolution MeerKAT and LOFAR images. In all case, we profiled the thermal and non-thermal emission using adjacent boxes of size equal to the resolution, and excluded any regions that lay on the diffuse tail of GB6~B0335+096.

\subsubsection{Surface brightness correlations}
The correlation between radio surface brightness $I_{\rm R}$ and X-ray surface brightness $I_{\rm X}$ provides insight into the physical processes at work in the ICM. The slope of the correlation provides insight into the underlying mechanism. In the case of the hadronic (secondary electron) model, a super-linear slope is generally expected due to the centrally-peaked distribution of cosmic ray protons (CRp) and how this scales with the thermal gas \citep[as discussed by][]{Ignesti2020_MHsample}. In the case of the turbulent (re-)acceleration scenario, either a sub-linear or super-linear slope can be recreated depending on the the distribution and properties of the CRe population in the ICM \citep[e.g.][]{Brunetti_Jones_2014,ZuHone2013,ZuHone2015}. Similarly, in a hybrid scenario \citep[][]{Brunetti_Blasi_2005,Brunetti_Lazarian_2011} whereby CRp and their secondaries undergo re-acceleration by turbulence, it is also possible to replicate either a sub- or super-linear slope \citep[see for example][]{Riseley2023_A1413}.

\begin{figure*}
    \begin{center}
    \includegraphics[width=0.475\textwidth]{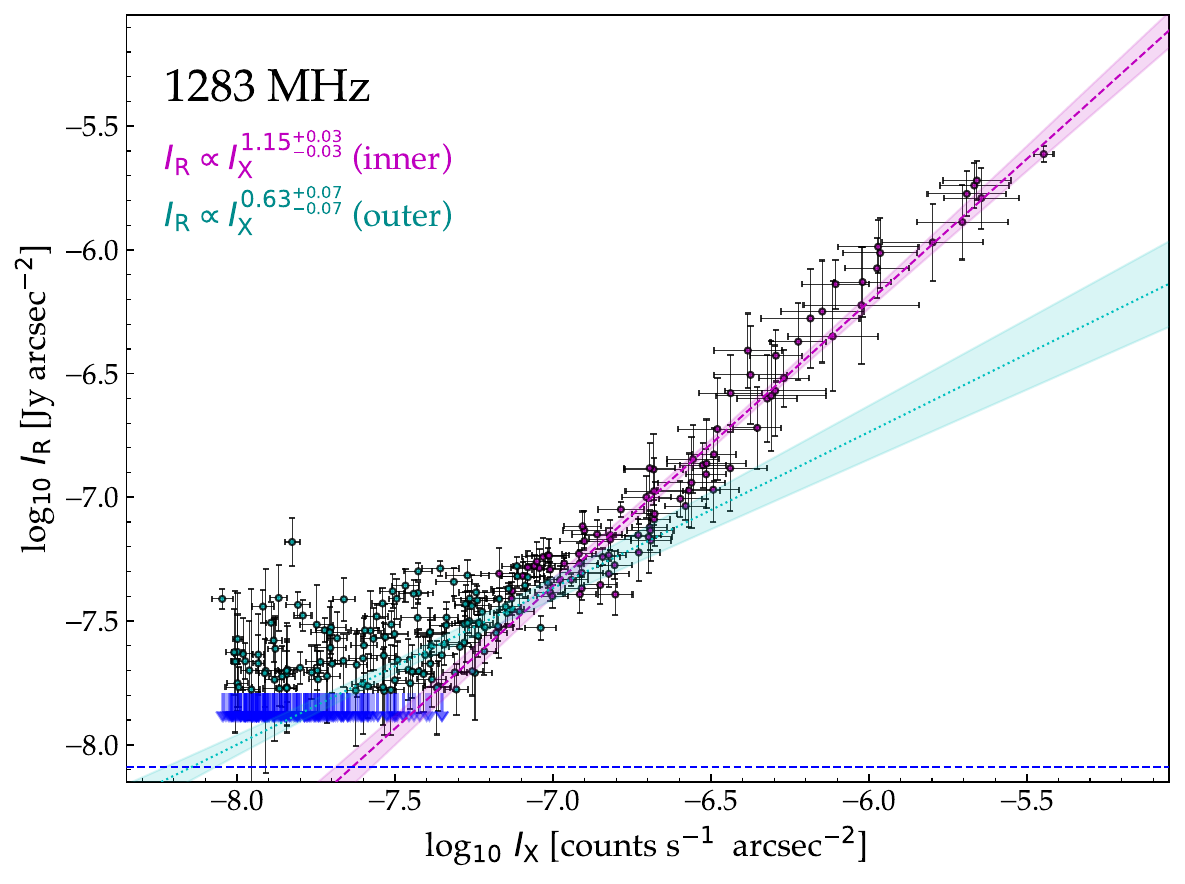}
    \includegraphics[width=0.475\textwidth]{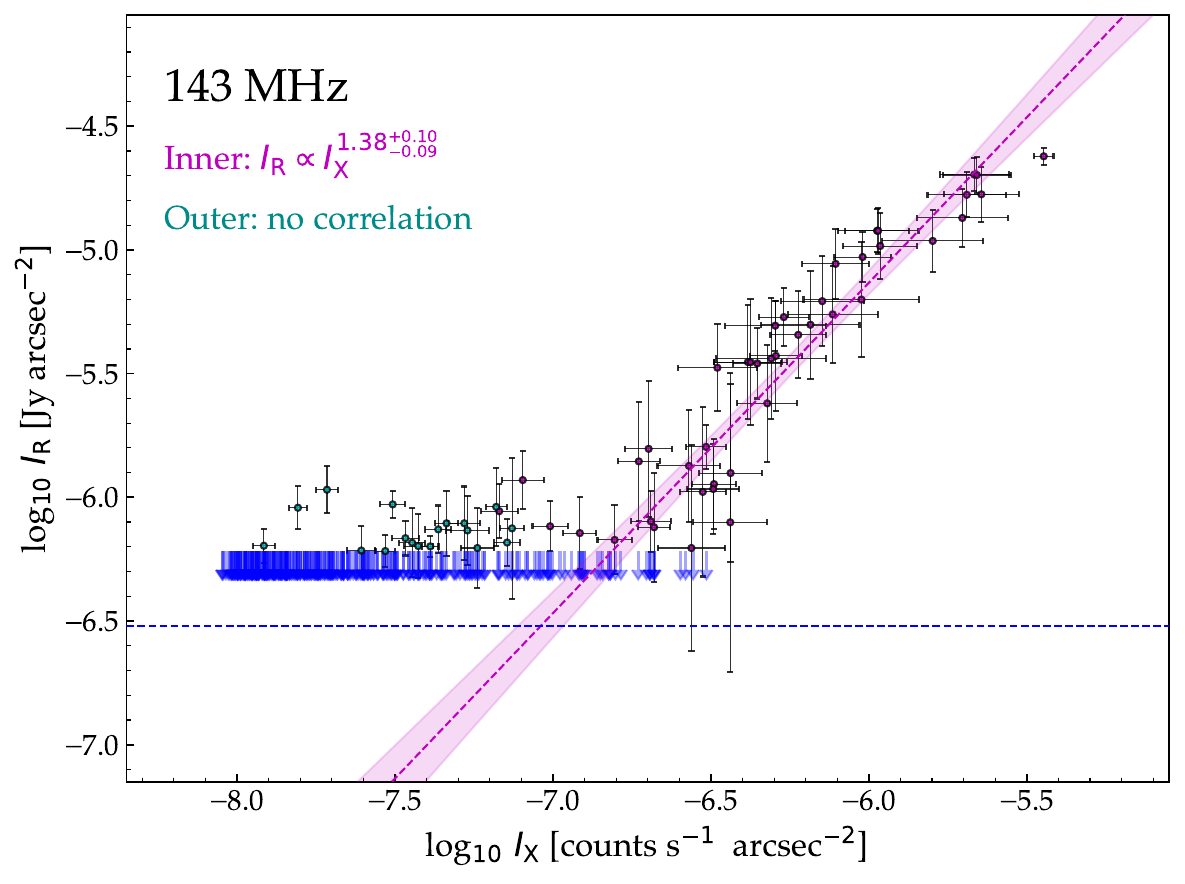}
    \caption{Point-to-point correlations between X-ray surface brightness $(I_X)$ and radio surface brightness $(I_R)$ for the diffuse multi-component halo in 2A~0335+096 at 1283\,MHz (left) and 145\,MHz (right) at 35\,arcsec resolution. The horizontal dashed line indicates the $1\sigma$ level, and blue arrows indicate $2\sigma$ upper limits.}
    \label{fig:2A0335_PTP_35asec}
    \end{center}
\end{figure*}

The point-to-point $I_{\rm R} / I_{\rm X}$ correlation at 35\,arcsec resolution is shown in Fig.~\ref{fig:2A0335_PTP_35asec}. It is immediately clear that there is departure from a single trend: above an X-ray surface brightness of $\log \left( I_{\rm X} \, [{\rm counts ~ s^{-1} ~ arcsec}^{-2}] \right) \gtrsim -7.0$ we see the data points are tightly correlated around a single slope, whereas below this threshold there is a far greater scatter in the data at a given X-ray surface brightness. 

As such, we characterised the relationship between $I_{\rm R}$ and $I_{\rm X}$ as following a double power-law in log-log space. Each power law follows the form:
\begin{equation}
    \log \left( I_{\rm R} \right) = b \log \left( I_{\rm X} \right) + c
\end{equation}
where the slope $b$ parametrises how the non-thermal emission scales with the thermal emission. We performed the fit using two independent sub-sets covering the inner and outer regions of the ``mini''-halo. To ensure robustness against an accidental ``sweet-spot'' boundary between inner and outer regions, we performed a Monte-Carlo analysis, which we detail in Appendix~\ref{appendix_mcmc}.

In brief, we use the clear break in behaviour seen in Fig.~\ref{fig:2A0335_PTP_T1D} at around $\log{I_{\rm X}} \simeq -7.2$  to define an initial boundary between inner and outer regions, which we then perturb to define differing sets of inner/outer regions. The initial set of inner/outer regions is shown in Fig.~\ref{fig:2A0335_boxes}. Each is used to perform the point-to-point analysis; we quote the median of all parameters as our final best-fit result in Table~\ref{tab:correlation_results} and plot this in Fig.~\ref{fig:2A0335_PTP_35asec}. See Appendix~\ref{appendix_mcmc} for full details.

\begin{table*}[]
    \centering
    \caption{Results for point-to-point correlations.}
    \label{tab:correlation_results}
    \renewcommand{\arraystretch}{1.25}
    \footnotesize
    \begin{tabular}{c c c c | c c | >{\raggedleft\arraybackslash}p{1.5cm} >{\raggedleft\arraybackslash}p{1.5cm}}
        Frequency & Resolution & $N_{\rm comp}$ & Correlation & Best-fit slope       & Intrinsic scatter      & Spearman coeff. & Pearson coeff. \\
                 & $[$arcsec$]$ &   &  & $b$       &  $\sigma_{\rm int}$     & $r_{\rm S}$ & $r_{\rm P}$ \\
        \hline\hline
        \multirow{2}{*}{1283\,MHz}  &  \multirow{2}{*}{35\,asec}  &  \multirow{2}{*}{2}  &   $I_R / I_X$ (inner)   &  ${1.15^{+0.03}_{-0.03}}$ &   $0.002^{+0.002}_{-0.001}$ & 0.95 &  ${{0.97}}$ \\
                &   &   &   $I_R / I_X$ (outer)   &  ${{0.63^{+0.07}_{-0.07}}}$ &   ${{0.039^{+0.008}_{-0.007}}}$ & ${0.41}$ &  ${0.39}$ \\
        \hline
        \multirow{2}{*}{145\,MHz}  &  \multirow{2}{*}{35\,asec}  &  \multirow{2}{*}{2}  &   $I_R / I_X$ (inner)   &  $1.38^{+0.10}_{-0.09}$ &   ${0.030^{+0.013}_{-0.008}}$ & ${0.94}$ &  ${0.92}$ \\
                &   &   &   $I_R / I_X$ (outer)   &  N/A &   N/A & ${0.10}$ &  ${0.09}$ \\
        \hline
        \multirow{2}{*}{1283\,MHz}  &  \multirow{2}{*}{35\,asec}  &  \multirow{2}{*}{2}  &   $\log(I_R) / T_{\rm X,1D}$ (inner)   &  $-1.09^{+0.05}_{-0.05}$ &   ${0.022^{+0.006}_{-0.005}}$ & ${-0.87}$ &  $-0.93$ \\
                &   &   &   $\log(I_R) / T_{\rm X,1D}$ (outer)   &  ${0.66^{+0.10}_{-0.11}}$ &   ${0.048^{+0.011}_{-0.009}}$ & ${0.33}$ &  ${0.29}$ \\
        \hline
        \multirow{2}{*}{N/A}  &  \multirow{2}{*}{35\,asec}  &  \multirow{2}{*}{2}  &   $\log(I_X) / T_{\rm X,1D}$ (inner)   &  ${-0.92^{+0.04}_{-0.04}}$ &   ${0.022^{+0.005}_{-0.004}}$ & ${-0.86}$ &  $-0.93$ \\
                &   &   &   $\log(I_X) / T_{\rm X,1D}$ (outer)   &  ${0.52^{+0.05}_{-0.05}}$ &   ${0.049^{+0.004}_{-0.004}}$ & 0.56 &  0.52 \\
        \hline
    \end{tabular}
\end{table*}

The fits were performed using the \texttt{Linmix} software package \citep{Kelly2007_LINMIX} which employs a Bayesian linear regression approach to determine the best-fit parameters. This incorporates uncertainties on both the independent and dependent variables, intrinsic scatter, and crucially upper limits on the dependent variable $(I_{\rm R})$. We used the MCMC implementation to derive the posterior distribution for the parameters, quoting the 50${\rm th}$ percentile as our best fit and the 16$^{\rm th}$ and 84$^{\rm th}$ percentiles as our $1\sigma$ uncertainties. We also determined the strength of the correlation using the Spearman and Pearson correlation coefficients, $r_{\rm S}$ and $r_{\rm P}$ respectively. Any regions which displayed a radio surface brightness below $2\sigma$ were treated as upper limits (blue arrows in Fig.~\ref{fig:2A0335_PTP_35asec}). The best-fit slope $b$, intrinsic scatter $\sigma_{\rm int}$ and correlation coefficients $r_{\rm S}$ and $r_{\rm P}$ are listed in Table~\ref{tab:correlation_results}, and our best-fit slopes are shown in Fig.~\ref{fig:2A0335_PTP_35asec}. 

For the inner component of the halo, we find a very strong correlation with nearly negligible intrinsic scatter at 1283\,MHz ($r_{\rm S} = 0.95$, ${r_{\rm P} = 0.97}$, $\sigma_{\rm int} = 0.002$) that follows a super-linear slope of ${b = 1.15^{+0.03}_{-0.03}}$. For the same component at 145\,MHz we see a similarly strong correlation (${r_{\rm S} = 0.94}$, ${r_{\rm P} = 0.92}$) with increased intrinsic scatter (${\sigma_{\rm int} = 0.030}$). This component follows an increasingly super-linear slope of $b = 1.38^{+0.10}_{-0.09}$ at 145\,MHz. These results are consistent with the coarser resolution point-to-point analysis from \cite{Ignesti2020_MHsample,Ignesti2021_2A0335} at 1.4\,GHz and 5.0\,GHz. However, our deeper analysis and improved source-subtraction allow us to both reliably remove contamination to a greater degree as well as study this correlation across the extent of the inner component (corresponding to the full extent of the mini-halo in previous studies) and so our results have significantly reduced uncertainties.

In the broader context of mini-haloes, the super-linear correlation slope at both frequencies is consistent with our previous results on MS~1455.0+2232 and Abell~1413 \citep[respectively][]{Riseley2022_MS1455,Riseley2023_A1413}. However, in contrast to these previous studies, in the case of 2A~0335+096 we see a steeper super-linear slope with LOFAR than with MeerKAT, whereas Abell~1413 demonstrated a steeper slope in the MeerKAT data at 1283\,MHz (suggesting spectral steepening with radius) and MS~1455.0+2232 showed a consistent slope at both frequencies (suggesting no radial steepening). The steeper slope at low frequencies seen for 2A~0335+096 might suggest a radial flattening of the spectral index toward higher frequencies, which is challenging to explain within the context of a purely turbulent (re-)acceleration scenario, but may suggest an increased importance of turbulent (re-)acceleration in the cluster outskirts compared to the inner region of the mini-halo.

For the outer component of the halo, the data show increased scatter at 1283\,MHz and a far weaker correlation (${r_{\rm S} = 0.41}$, ${r_{\rm P} = 0.39}$, ${\sigma_{\rm int} = 0.039}$) with a sub-linear slope of ${b = 0.63^{+0.07}_{-0.07}}$; at 145\,MHz our analysis shows negligible correlation between radio and X-ray surface brightness. This is likely due to the poor sensitivity of the LOFAR data compared to MeerKAT in the outer regions of the halo, as the vast majority of regions provide only upper limits.

\subsubsection{X-ray temperature correlations}
Given the high quality of our \textit{XMM-Newton} data, we were also able to investigate correlations between both radio and X-ray surface brightness and X-ray temperature. For each region, we derived 1D X-ray temperatures from our 2D X-ray temperature and surface brightness maps using equations (2) and (3) of \cite{Lovisari2024_CHEX-MATE}. We refer the reader to that paper for the detailed methodology and background.

\begin{figure*}
    \begin{center}
    \includegraphics[width=0.475\textwidth]{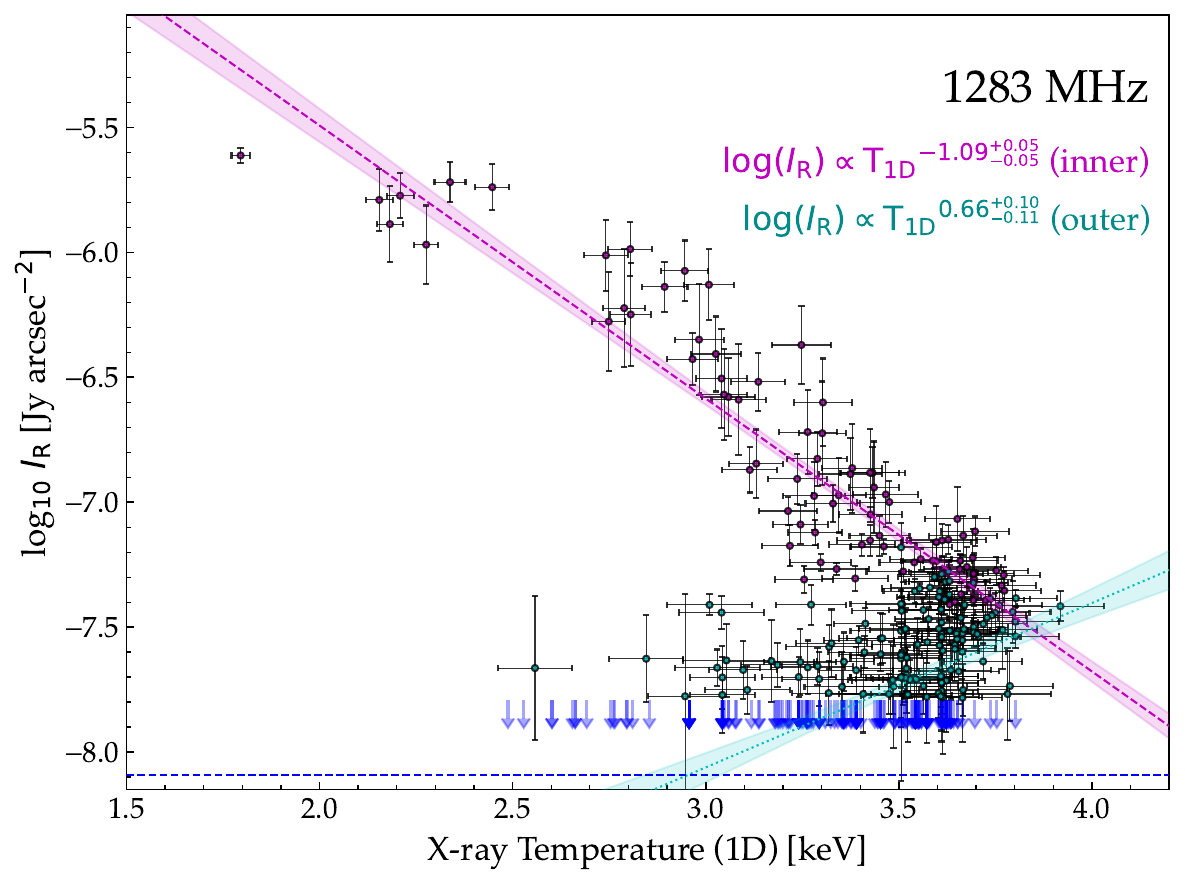}
    \includegraphics[width=0.475\textwidth]{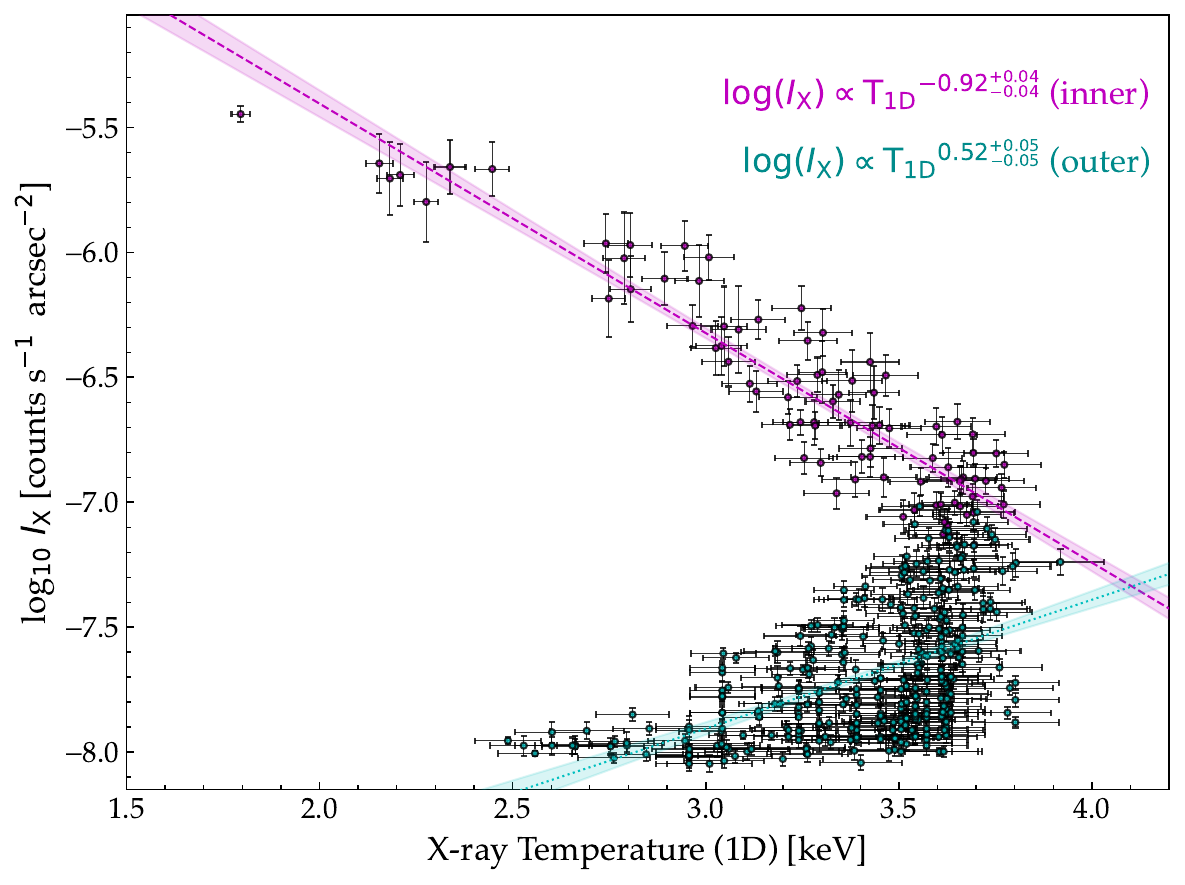}
    \caption{Spatial correlation between 1D X-ray temperature $T_{\rm X, 1D}$ and radio surface brightness $\log(I_R)$ at 1283\,MHz (left) and X-ray surface brightness $\log(I_X)$ (right) for the diffuse multi-component halo in 2A~0335+096 at 35\,arcsec resolution. The horizontal dashed line indicates the $1\sigma$ level, and blue arrows indicate $2\sigma$ upper limits.}
    \label{fig:2A0335_PTP_T1D}
    \end{center}
\end{figure*}

We parameterised the relationship between X-ray temperature $T_{\rm X, 1D}$ and radio surface brightness or X-ray surface brightness using a power-law form in log-linear space as:
\begin{equation}
    \log \left( I \right) = b \cdot T_{\rm X, 1D}  + c
\end{equation}

The correlations between 1D X-ray temperature and radio surface brightness, $ \log(I_{\rm R})/T_{\rm X, 1D}$ and X-ray surface brightness, $\log(I_{\rm X})/T_{\rm X, 1D}$ are shown in Fig.~\ref{fig:2A0335_PTP_T1D}.

Our results show that for the inner component, there is an anti-correlation between the X-ray temperature and both the X-ray and radio surface brightness, i.e. colder regions show increased radio and X-ray surface brightness. The correlation coefficients are strong (${r_{\rm S} = -0.86,\, -0.87}$, $r_{\rm P} = -0.93$) and the intrinsic scatter is relatively low (${\sigma_{\rm int} = 0.022}$). The slope of the correlation is steeper for the $\log(I_{\rm R})/T_{\rm X, 1D}$ correlation ($b = -1.09^{+0.05}_{-0.05}$) than for the $\log(I_{\rm X})/T_{\rm X, 1D}$ correlation (${b = -0.92^{+0.04}_{-0.04}}$), implying that the radio surface brightness increases more steeply with decreasing X-ray temperature than X-ray surface brightness.

For the outer component, the correlation coefficients are far weaker. While they are indicative of a weak-to-moderate correlation, suggesting that hotter regions show an increased radio and X-ray surface brightness, the intrinsic scatter is higher. We caution against over-interpretation of this result however: the signal-to-noise requirements result in an increased region size further from the cluster centre, meaning that many of the cyan regions may be drawn from the same 2D temperature cell(s). Additionally, projection effects play a strong role.

\subsection{Radial profiles}
The strong correlation between the radio and X-ray point-to-point correlation for the inner component implies a strong connection between the non-thermal and thermal components of the ICM, although the super-linear slope indicates that the radio surface brightness decreases more rapidly than the X-ray surface brightness towards larger radii. 

\begin{figure}
    \begin{center}
    \includegraphics[width=0.45\textwidth]{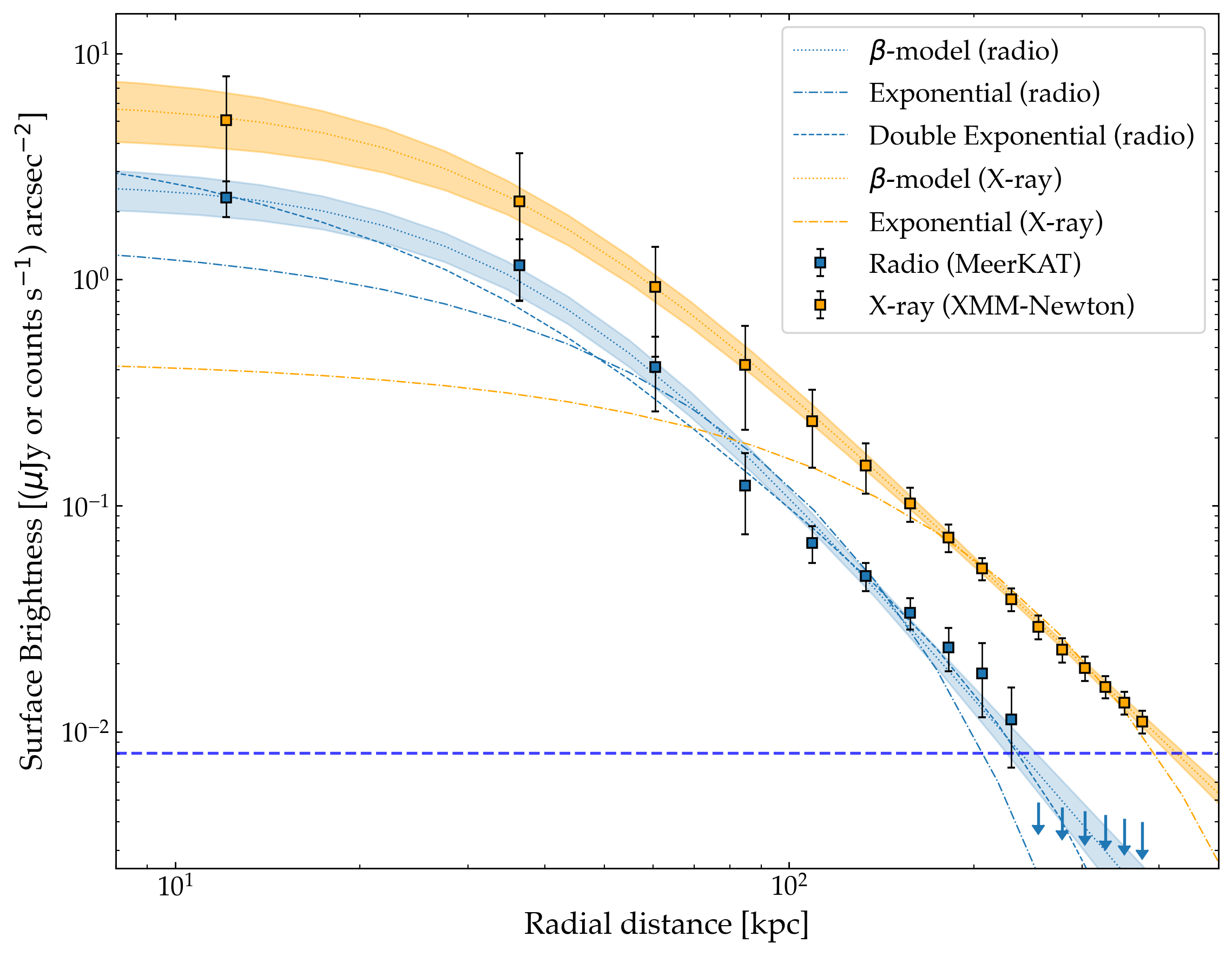}
    \caption{Radial profiles of the non-thermal (traced by MeerKAT) and thermal (traced by \textit{XMM-Newton}) surface brightness. Dotted, dashed and dot-dashed lines trace different model fits to the data, with the shaded region denoting the $1\sigma$ uncertainty. For clarity we show this uncertainty for the $\beta$-model only, which provides the best descriptor of our data. The horizontal line indicates the $1\sigma$ level for our MeerKAT image.}
    \label{fig:2A0335_radial_profile}
    \vspace{-3mm}
    \end{center}
\end{figure}

We explored this further by deriving radial profiles of both the radio and X-ray surface brightness, as traced by MeerKAT and \textit{XMM-Newton} at 35\,arcsec resolution. These profiles are shown in Fig.~\ref{fig:2A0335_radial_profile}, with the radial bins used shown in Fig.~\ref{fig:2A0335_wedges}. As expected from the point-to-point correlation, we see a good correspondence between the two profiles. 

Radio (mini-)haloes are often modelled using single- or double-exponential profiles to describe the non-thermal component \citep[e.g.][]{Murgia2009,Boxelaar2021_HaloFDCA,Riseley2022_MS1455,Riseley2022_A3266,vanWeeren2026_A1775_A1795} and a single- or double- form of the $\beta$-model \citep{Cavaliere_FuscoFemiano_1976,Arnaud2009} for the thermal component \citep[e.g.][]{Pasini2019_A2495,Ubertosi2021_A795,Pasini2021_A1668,Lusetti2024_A1413}. 

As such we attempted to characterise our radial profile at 1283\,MHz using both single- and double-exponential functions of the form:
\begin{equation}
    I(r) = \sum_{n=1}^{2} I_n \exp(-r/r_{e,n})
\end{equation}
where $I_{c,n}$ is the central surface brightness for component $n$ and $r_{e,n}$ is the $e$-folding radius. For the thermal surface brightness profile, we attempted to fit a single exponential as well as the $\beta$-model. This model takes the following form:
\begin{equation}
    I(r) = I_c \left[ 1 + \left( \frac{r}{r_c} \right)^2 \right]^{\left(\frac{1}{2} - 3\beta\right) }
\end{equation}
where $I_{c}$ is the central surface brightness, $r_c$ is the core radius, and $\beta$ describes the ratio between thermal and gravitational energy, essentially affecting how rapidly the surface brightness decreases with radius. The best-fit parameters for each model are listed in Table~\ref{tab:profile_fits} and presented in Fig.~\ref{fig:2A0335_radial_profile}. However, it is clear that both the single- and double-exponential model fail to replicate the observed surface brightness profile of both the non-thermal and thermal emission; as such, we also fitted a $\beta$-model to our non-thermal profile. This fit is similarly shown in Fig.~\ref{fig:2A0335_radial_profile}; the $\beta$-model provides a very good reproduction of both the non-thermal and thermal surface brightness profiles, as also found by \cite{vanWeeren2026_A1775_A1795} for Abell~1795.

The best-fit profiles show consistent core radii $r_c$ in both the non-thermal and thermal, respectively $39.5^{+7.5}_{-6.4}$\,kpc and $33.5^{+7.8}_{-6.1}$\,kpc. This radius lies within the extent of inner component---the classical mini-halo---which spans a radius between 95 and 135\,arcsec (65 to 93\,kpc) from the same reference point.

The slope of the profile is steeper for the non-thermal than the thermal, as we find $\beta_{\rm R} = 0.70^{+0.06}_{-0.05}$ for the radio fit and $\beta_{\rm X} = 0.59^{+0.02}_{-0.02}$ for the X-ray fit, indicating that the radio surface brightness decreases more rapidly as a function of radius than the X-ray surface brightness, consistent with the super-linear slope in the point-to-point correlations. However, despite the evidence of two distinct behaviours in the point-to-point correlations, we do not see evidence for departure from a single $\beta$-model in either the thermal or non-thermal profiles.

\begin{table*}[]
    \centering
    \caption{Fit parameters for radial profiles.}
    \label{tab:profile_fits}
    \renewcommand{\arraystretch}{1.25}
    \footnotesize
    \begin{tabular}{l l | c c c c | c c c }
                &       & \multicolumn{4}{c |}{Double-exponential fit} & \multicolumn{3}{c}{$\beta$-model fit} \\
        Profile & Model & $I_{c,1}$   &   $r_{e,1}$     &   $I_{c,2}$   &   $r_{e,2}$     &  $I_c$  &  $r_c$ &  $\beta$ \\
        \hline\hline
        \multirow{3}{*}{Radio} &    Single exponential &   $1.57^{+0.60}_{-0.48}$  & $39.3^{+5.1}_{-3.8}$ & $-$ & $-$ & $-$ & $-$ & $-$ \\
        &   Double exponential &   $4.12^{+1.22}_{-1.01}$  & $17.7^{+4.0}_{-4.4}$ & $0.43^{+0.18}_{-0.15}$  & $59.3^{+10.4}_{-7.2}$ & $-$ & $-$ & $-$\\
        &   $\beta$-model &   $-$ & $-$ & $-$ & $-$ & $2.69^{+0.57}_{-0.55}$ & $39.5^{+7.5}_{-6.4}$ & $0.70^{+0.06}_{-0.05}$ \\
        \hline
        \multirow{2}{*}{X-ray} &    Single exponential &   $0.45^{+0.09}_{-0.07}$  & $98.1^{+6.3}_{-5.6}$ & $-$ & $-$ & $-$ & $-$ & $-$ \\
        &   $\beta$-model &   $-$ & $-$ & $-$ & $-$ & $5.41^{+2.48}_{-1.78}$ & $33.5^{+7.8}_{-6.1}$ & $0.59^{+0.02}_{-0.02}$ \\
        \hline
    \end{tabular}
    \tablefoot{
    For the double-exponential fits, $I_c$ is the core surface brightness for the respective component and $r_e$ is the e-folding radius. For the $\beta$-model fit, $I_c$ is the core surface brightness, $r_c$ is the core radius and $\beta$ is the beta parameter.
    }
\end{table*}

\subsection{Radio/X-ray residuals}
Given the close correspondence between the thermal and non-thermal correlations and the radial profiles reported in the previous sections, following \cite{vanWeeren2026_A1775_A1795} we calculated a radio/X-ray residual (RXR) in order to search for evidence of deviation(s) from the calculated point-to-point correlations. Here we used our MeerKAT for the radio data and our \textit{XMM-Newton} image for the X-ray data, both at 35\,arcsec resolution.

We computed the RXR as:
\begin{equation}
    \label{eq:rxr}
    {\rm RXR} = I_{\rm R} - 10^{c} \cdot I_{\rm X}^b 
\end{equation}
where $I_{\rm R}$ and $I_{\rm X}$ are the radio and X-ray images, respectively, normalised to the central brightness $I_c$ from the respective profiles, and $b$ and $c$ are obtained from the point-to-point correlation. In this case, we computed the RXR using the point-to-point fitting results for the inner component of the halo derived at 35\,arcsec resolution. We blanked pixels associated with the head-tail galaxy GB6~B0335+096, for clarity.

\begin{figure}
    \begin{center}
    \includegraphics[width=0.45\textwidth]{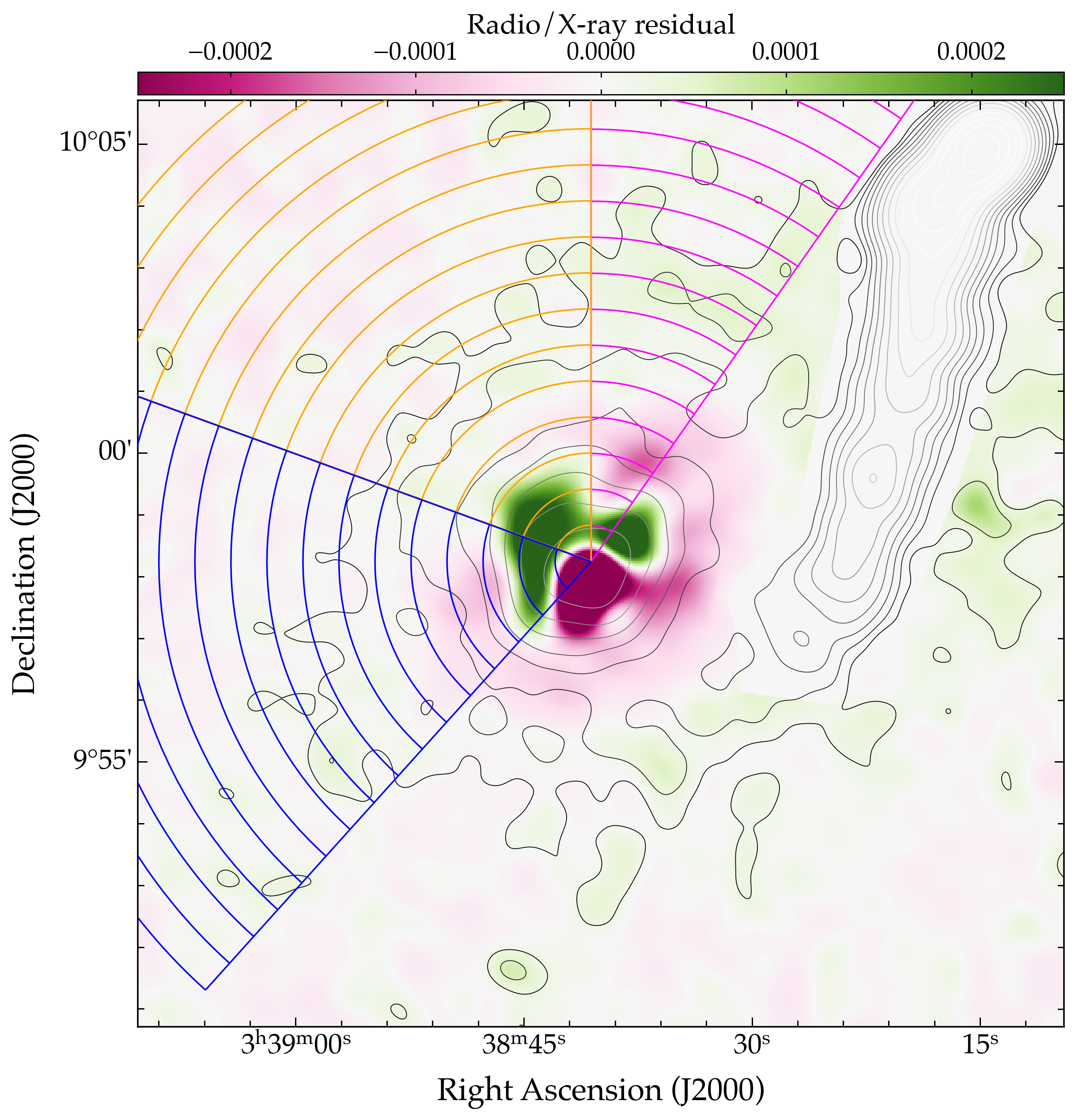}
    \caption{Radio/X-ray residual image of 2A~0335+096 derived using Equation~\ref{eq:rxr}. For clarity, the head-tail galaxy GB6~B0335+096 has been blanked. Contours show the MeerKAT image at 35\,arcsec resolution starting at $3\sigma$ and scaling by a factor $2$. Green colours indicate an excess of radio emission; magenta colours indicate an excess of X-ray emission. Wedges mark sectors used to derive additional profiles shown later.}
    \label{fig:2A0335_RXR}
    \end{center}
\end{figure}

Fig.~\ref{fig:2A0335_RXR} shows the derived RXR, which reveals several features of significance: two inner regions of radio excess close to the cluster centre, and one outer region to the north-west where the larger halo fills out to the tailed radio galaxy GB6~B0335+096. The inner radio excess regions have different origins: the north-western excess is co-located with the X-ray cavity seen by \textit{Chandra}, whereas the broader north-eastern excess is co-located with the extended north-eastern `wing' of the mini-halo seen at 15\,arcsec resolution (see Fig.~\ref{fig:2A0335_diffuse_15asec}).

This tentative excess motivates us to examine radial profiles along different sectors as an independent check. We split our previous radial profile, which covered position angles from $-35^\circ$ to $138^\circ$, into three sectors covering $-35^\circ$ to $0^\circ$ (NW), $0^\circ$ to $70^\circ$ (NE) and $70^\circ$ to $138^\circ$ (SE). These sectors are shown in Fig.~\ref{fig:2A0335_RXR}; Fig.~\ref{fig:2A0335_radial_profile_arcs} shows the profiles along this sectors, along with the residuals compared to the best-fit global $\beta$-model from earlier.

For the NW arc, which traces the region of the radio excess in our RXR, Fig.~\ref{fig:2A0335_radial_profile_arcs} confirms a statistically-significant excess of up to a factor $\sim2$ compared to our global $\beta$-model fit beyond a radius of around 100\,kpc. This excess traces the emergence of the larger-scale radio halo which fills the region between the classical mini-halo and the head-tail radio galaxy GB6~B0335+096.

\section{Discussion: on the nature of the halo in 2A~0335+096}

Our new deep MeerKAT observations have enabled the first detection of a large-scale ``mini''-halo spanning over 500\,kpc in 2A~0335+096. This diffuse radio source has an integrated flux density of $34.8 \pm 1.8$\,mJy at 1283\,MHz, corresponding to a $k$-corrected radio power of $P_{\rm 1.4\,GHz} = (8.86 \pm 0.46) \times 10^{22}$~W\,Hz$^{-1}$. With a cluster mass $M_{500} = 2.27 \times 10^{14}~{\rm M_{\odot}}$ \citep{Planck2014} this places 2A~0335+096 toward the low-mass end of the scaling relation between radio luminosity and cluster mass for mini-halo hosting clusters \citep[see][and references therein]{Kolokythas2025_MGCLS2}.

This emission is complex, and would appear to show a more classical mini-halo coexisting with a larger-scale halo, as seen in some relaxed clusters \citep[see for example][and references therein]{vanWeeren2026_A1775_A1795}. However, verifying the existence of a multi-component halo is challenging and requires careful consideration of the evidence as well as accurate source subtraction \citep[see discussion in][]{Rajpurohit2025_RadioHalos}. We will now consider our findings.

From our point-to-point correlations, we see evidence of two different trends: the inner region of the halo demonstrates a strong thermal/non-thermal connection in the $I_R/I_X$ plane with a super-linear slope, whereas the outer region of the halo exhibits a sub-linear slope, although the correlation strength is weaker. Similarly, we see different trends in the $I_R/T_{\rm 1D}$ in these regions: the inner part of the halo shows a strong anticorrelation (cooler regions are brighter) whereas the outer part shows a weak correlation (warmer regions are brighter) although we caution against interpreting this outer trend too strongly.

In contrast, the radial profiles are generally well-described by only a single trend in both the radio and X-ray surface brightness. Both are best described by a $\beta$-model, supporting the interpretation of a close correspondence between the thermal and non-thermal components. However, examining profiles along different sectors---motivated by the radio excess seen in some regions of our RXR image---confirms an excess of radio emission (with respect to the global $\beta$-model) towards the north-west, in the direction of the tailed radio galaxy GB6~B0335+096.

The halo shows a generally steep spectral index, with an average spectral index of $\langle \alpha \rangle = -1.20 \pm 0.09$, steeper than typically expected for a hadronic scenario. Our results show a strong correlation between thermal and non-thermal components, both of which are well-fit by a single $\beta$-model when examining the radial profiles. The complexity of the halo, as inferred from the presence of multiple components in the point-to-point correlation, arises more naturally within a turbulent acceleration framework; both the super-linear slope found for the inner component and the sub-linear slope found for the outer component are replicable within this framework, although the super-linear slope of the inner component is likewise produced in a hadronic scenario. Similarly, the lack of clear radial steepening in the spectral index is more in line with the hadronic scenario than the turbulent acceleration framework.

2A~0335+096 hosts one reported cold front around 40\,kpc south of the cluster centre \citep{Mazzotta2003_2A0335,Sanders2009_2A0335} suggesting sloshing within the cluster potential well. The presence of a larger-scale halo in 2A~0335+096 is thus in line with the growing trend of cluster-scale diffuse emission in sloshing cool-core clusters \citep[e.g.][]{Biava2024_sloshing_MH}.

Since 2A~0335+096 is a relaxed cluster however, within a turbulent (re-)acceleration framework it would be expected that the outer halo component would exhibit a steeper spectrum than the inner component due to lower levels of turbulent energy. Such behaviour is seen---albeit on far larger scales---in radio mega-haloes \citep{Cuciti2022_megahaloes}. Our results are inconsistent with such a scenario, as the outer halo component is not detected by LOFAR; moreover the spectral index is more consistent with the spectra of giant radio haloes in massive merging clusters \citep[see for example][and references therein]{Rajpurohit2025_RadioHalos}. One potential explanation could be some level of disturbance in the ICM beyond the cool core region, although our data do not allow us to investigate this in detail.

Regarding the hadronic scenario: in 2A~0335+096 there are two clear sources of CRp injection into the central mini-halo, namely the BCG---which through its radio-bright core and multiple pairs of lobes has been demonstrably active for a significant period---and the head-tail galaxy GB6~B0335+096. While diffusion of CRp over the full extent of the larger-scale halo is not viable, numerical modelling suggests CRp diffusion coefficients of the order of $\sim(1 - 10)\times10^{29}~{\rm cm^2 \, s^{-1}}$ in cluster environments, suggesting diffusion scales of up to $\sim 100~{\rm kpc}$ \citep[see for example][]{Brunetti_Jones_2014,Ignesti2020_MHsample}. This length scale is fully compatible with the extent of the inner component, the classical mini-halo in 2A~0335+096. 

As such, within the context of determining the nature of the mechanism powering the mini-halo, our evidence suggests a mixed picture. Neither mechanism is more clearly able to reproduce all of our findings, although a turbulent (re-)acceleration interpretation may be marginally favoured. Instead, we may be seeing an increasingly important contribution from hadronic processes---leading to a hybrid hadronic/turbulence mechanism---toward the cluster centre, while the larger-scale halo emission is likely turbulence-dominated.

\section{Conclusions}
We have presented new, deep MeerKAT L-band (1283\,MHz) and LOFAR 145\,MHz continuum observations of the sloshing cool core cluster 2A~0335+096. Our data provide a wealth of information on the radio source population of this cluster, from the known mini-halo, to the extended radio galaxies in the cluster and viewed in projection onto the ICM. In this paper we focussed on the diffuse emission; a full spectropolarimetric analysis of the radio galaxies will be presented in a further paper.

Our study has revealed the presence of a cluster-scale diffuse radio halo which extends for around 510\,kpc. This extended halo is centred on the known mini-halo but is asymmetric, extending further to the north/north-west than the south/south-east. Only the known mini-halo is detected by LOFAR, but lower limits on the radio spectrum of the extended halo from the LOFAR non-detection are consistent with the average spectral index of the mini-halo, which we find to be $\alpha = -1.20 \pm 0.09$ from our resolved analysis. 

We quantified the correlations between radio- and X-ray surface brightness ($I_{\rm R}/I_{\rm X}$), radio surface brightness and X-ray temperature ($\log(I_{\rm R})/T_{\rm X,1D}$) and X-ray surface brightness and X-ray temperature ($\log(I_{\rm X})/T_{\rm X,1D}$). We found evidence of two distinct behaviours in different regions of the cluster: the central component of the halo shows a strong correlation with a super-linear slope in the $I_{\rm R}/I_{\rm X}$ plane at both 1283\,MHz and 145\,MHz. The outer component can only be explored at 1283\,MHz but shows a sub-linear slope with a moderate correlation. We verified that this result is robust against an accidental ``sweet spot'' division between inner and outer components via a Monte-Carlo analysis. We see a strong anti-correlation in the $\log(I_{\rm R})/T_{\rm X,1D}$ plane for the central component, but only a mild correlation for the outer component. 

We constructed radial profiles for both the thermal and non-thermal components, finding that both are best described by a single $\beta$-model. The core radii are similar but the $\beta$-parameter is larger for the radio profile than for the X-ray profile, implying that the non-thermal surface brightness decreases more rapidly with radius, consistent with the super-linear slope in the point-to-point correlation. 

We derived a radio/X-ray residual image which reveals three features of interest. We find two radio excess features close to the cluster centre: the first corresponds to the known X-ray cavity in the ICM, which is driven by AGN feedback from the BCG, whereas the second corresponds to the north-eastern extension of the mini-halo seen in particular by MeerKAT at 15\,arcsec resolution. The third radio excess lies to the north-west and corresponds to the large extension of the radio halo, which fills the cluster volume out to the location of GB6~B0335+096. Motivated by this contiguous large-scale radio excess we derived radial profiles in three sectors; the north-western sector profile confirms the presence of this excess over a single $\beta$-model.

We evaluated our results in the context of attempting to determine between a hadronic or turbulent (re-)acceleration mechanism powering the diffuse emission. Our findings do not allow us to clearly discriminate between mechanisms, although a turbulent (re-)acceleration scenario may be slightly favoured. Instead, we may be seeing a hybrid mechanism at work with the outer regions of the halo turbulence-dominated, but with an increasingly important contribution from hadronic processes toward the cluster centre, dominated by CRp injected by the central BCG.

Overall our results continue to support the emerging trend of cluster-scale diffuse emission in sloshing cool-core clusters, blurring the once-distinct classifications of radio haloes/mini-haloes. This indicates that deep, high-dynamic-range multi-frequency observations are highly motivated to complete our understanding of radio halo characteristics and structure in connection to cluster dynamics.

\begin{acknowledgement}
We thank our referee for their positive and insightful comments on our paper, which have improved the robustness of our analysis.
CJR acknowledges financial support from the German Science Foundation DFG, via the Collaborative Research Center SFB1491 `Cosmic Interacting Matters – From Source to Signal', as well as from the German Federal Ministry of Education and Research (BMFTR) under grant 05A26PCA (Verbundprojekt LOFAR-2-SKA).
LL acknowledges support from INAF grant 1.05.12.04.01. 
This research made use of the LOFAR-IT computing infrastructure supported and operated by INAF, including the resources within the PLEIADI special ``LOFAR'' project by USC-C of INAF, and by the Physics Dept. of Turin University (under the agreement with Consorzio Interuniversitario per la Fisica Spaziale) at the C3S Supercomputing Centre, Italy.
RT is grateful for support from the UKRI Future Leaders Fellowship (grant MR/Y020405/1). This work was supported by the STFC [grants ST/T000244/1, ST/V002406/1].
ABonafede acknowledges support from the ERC CoG $\vec{B}ELOVED$, GA n. GA N.101169773.
AI acknowledges support from the institutional project RVO:67985815 and the project 25-19512L of the Czech Science Foundation. 
NB acknowledges support from the ERC Consolidator Grant ULU 101086378.

The MeerKAT telescope is operated by the South African Radio Astronomy Observatory, which is a facility of the National Research Foundation, an agency of the Department of Science and Innovation. We wish to acknowledge the assistance of the MeerKAT science operations team in both preparing for and executing the observations used in this paper.

LOFAR is the Low Frequency Array designed and constructed by ASTRON. It has observing, data processing, and data storage facilities in several countries, which are owned by various parties (each with their own funding sources), and which are collectively operated by the ILT foundation under a joint scientific policy. The ILT resources have benefited from the following recent major funding sources: CNRS-INSU, Observatoire de Paris and Universit\'e d'Orl\'eans, France; BMBF, MIWF-NRW, MPG, Germany; Science Foundation Ireland (SFI), Department of Business, Enterprise and Innovation (DBEI), Ireland; NWO, The Netherlands; The Science and Technology Facilities Council, UK; Ministry of Science and Higher Education, Poland; The Istituto Nazionale di Astrofisica (INAF), Italy.

This research made use of the Dutch national e-infrastructure with support of the SURF Cooperative (e-infra 180169) and the LOFAR e-infra group. The J\"{u}lich LOFAR Long Term Archive and the German LOFAR network are both coordinated and operated by the J\"{u}lich Supercomputing Centre (JSC), and computing resources on the supercomputer JUWELS at JSC were provided by the Gauss Centre for Supercomputing e.V. (grant CHTB00) through the John von Neumann Institute for Computing (NIC).

This research made use of the University of Hertfordshire high-performance computing facility and the LOFAR-UK computing facility located at the University of Hertfordshire and supported by STFC [ST/P000096/1], and of the Italian LOFAR IT computing infrastructure supported and operated by INAF, and by the Physics Department of Turin university (under an agreement with Consorzio Interuniversitario per la Fisica Spaziale) at the C3S Supercomputing Centre, Italy.

This research has also made use of the following python packages not specifically cited elsewhere in the text: \textsc{astropy}, a community-developed core Python package for Astronomy \citep{Astropy_2013,Astropy_2018,Astropy_2022}, \textsc{aplpy} \citep{Robitaille2012}, \textsc{cmasher} \citep{vanderVelden2020}, \textsc{colorcet} \citep{Kovesi2015}, \textsc{matplotlib} \citep{Hunter2007}, \textsc{numpy} \citep{Numpy2011,Harris_2020_NumPy} and \textsc{scipy} \citep{Jones2001}. This research made extensive use of the Astrophysics Data System (ADS), funded by NASA under Cooperative Agreement 80NSSC21M00561.

\end{acknowledgement}

\vspace{-5mm}

\bibliographystyle{aa}
\bibliography{aa61262-26}

\clearpage

\fancypagestyle{appstyle}{
\fancyhf{}
\chead{C. J. Riseley et al.: MeerKAT-meets-LOFAR: 2A~0335}
\rfoot{aa61262-26, A1}
}

\pagestyle{appstyle}

\onecolumn

\begin{appendix}
\section{Wide-field images of 2A~0335+096}\label{appendix_widefield}

\begin{figure*}
    \sidecaption
    \includegraphics[width=12cm]{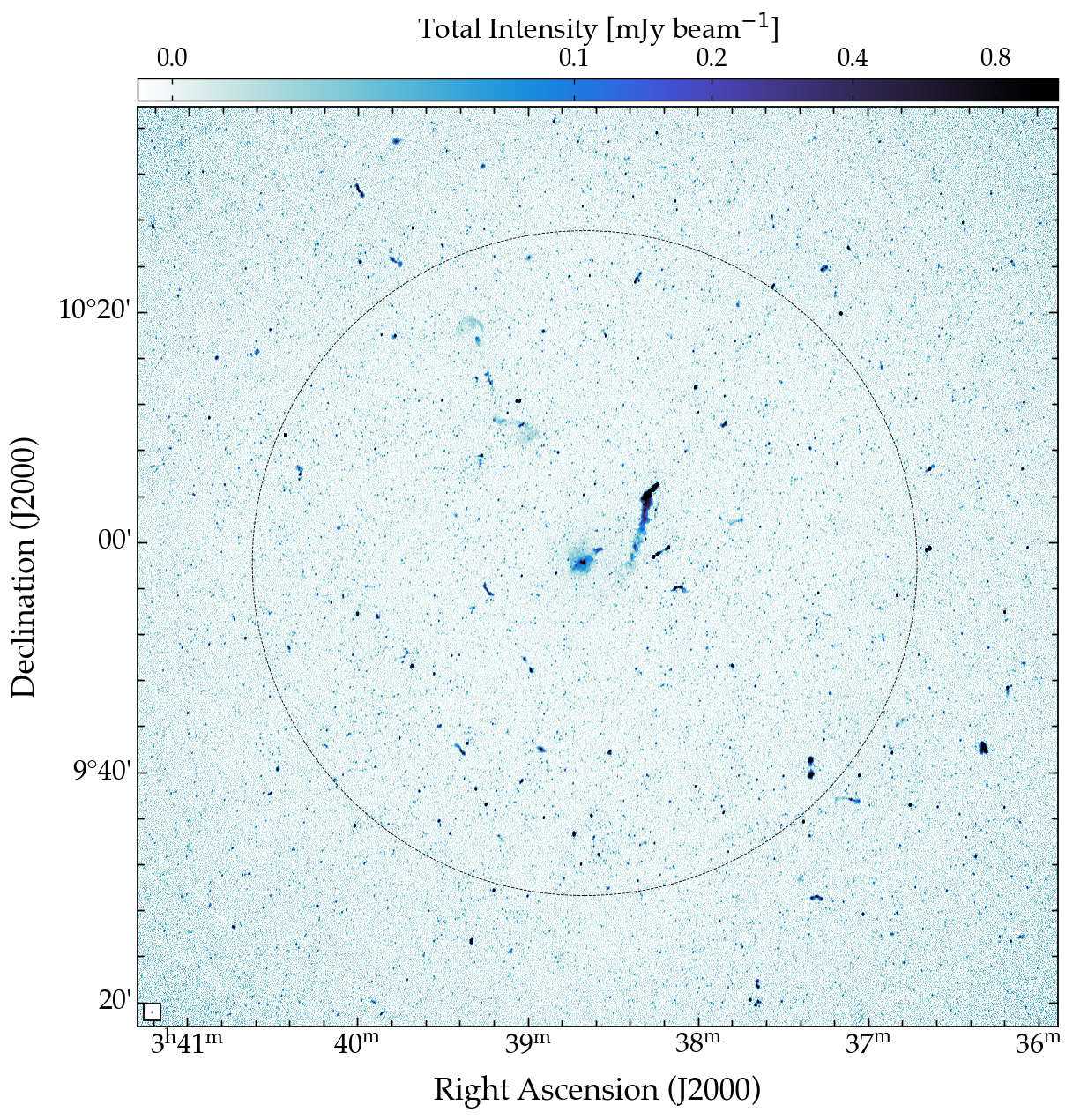}
    \caption{Full-field MeerKAT image of 2A~0335+096 at 1283\,MHz. The off-source noise is $\sigma = 4.3~\upmu{\rm Jy ~ beam}^{-1}$ where the restoring beam is 6.8\,arcsec by 4.7\,arcsec at a PA of 170\,deg. The colour scale runs from $-1\sigma$ to $250\sigma$ on an arcsinh stretch. The dashed circle denotes a radius of 28.9~arcminutes, corresponding to a physical scale of 1.2~Mpc at the redshift of 2A~0335+096 ($z = 0.0363$; \citealt{Sanders2011_clusters_XMM}).}
    \label{fig:2A0335_fullfield}
\end{figure*}

We show our wide-field primary beam corrected MeerKAT image of 2A~0335+096 in Fig.~\ref{fig:2A0335_fullfield}.

\section{Regions for exploration of thermal and non-thermal correlations and profiles}\label{appendix_regions}

\begin{figure}[!hbp]
    \begin{center}
    \includegraphics[width=0.9\textwidth]{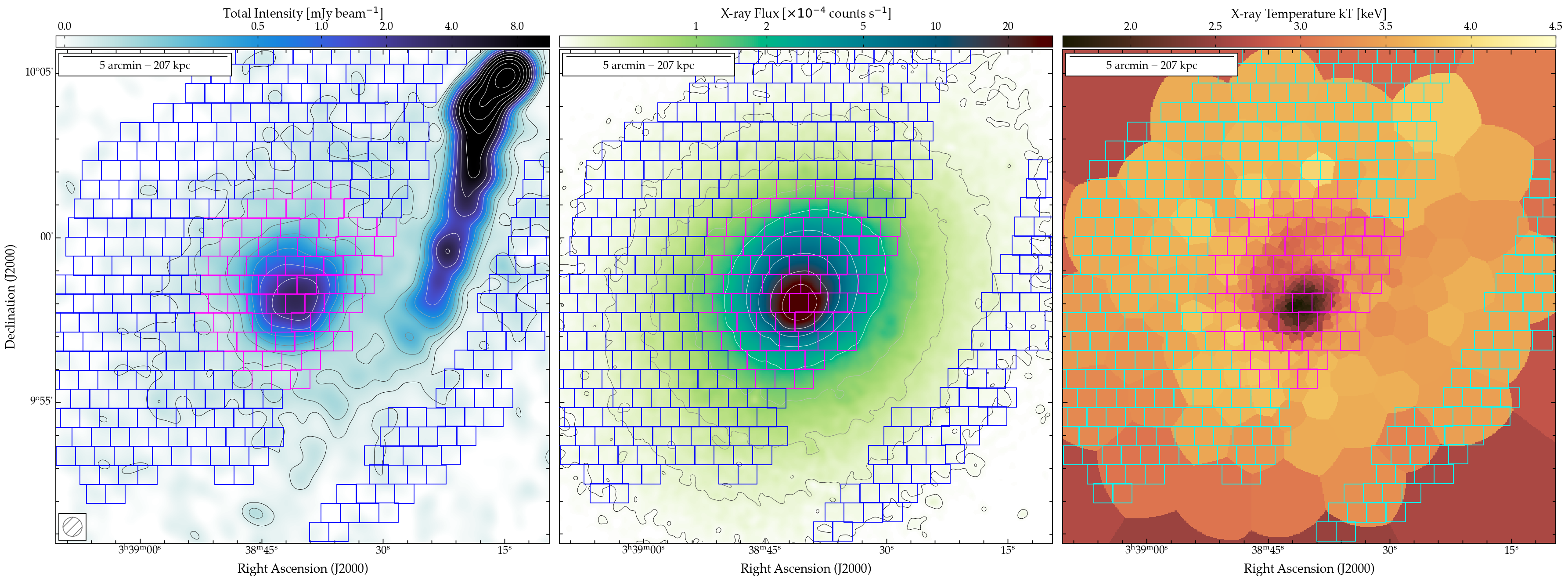}
    \caption{Images of 2A~0335+096 and the associated initial region sets used for characterising the point-to-point correlations. The MeerKAT image (left) is as per Fig.~\ref{fig:2A0335_diffuse_35asec}; the \textit{XMM-Newton} surface brightness (centre) and temperature (right) images are as per Fig.~\ref{fig:2A0335_cluster_presub_xray}. For clarity, contours scale by a factor 2 rather than $\sqrt{2}$.}
    \label{fig:2A0335_boxes}
    \vspace{-3mm}
    \end{center}
\end{figure}

\begin{figure*}
    \sidecaption
    \includegraphics[width=12cm]{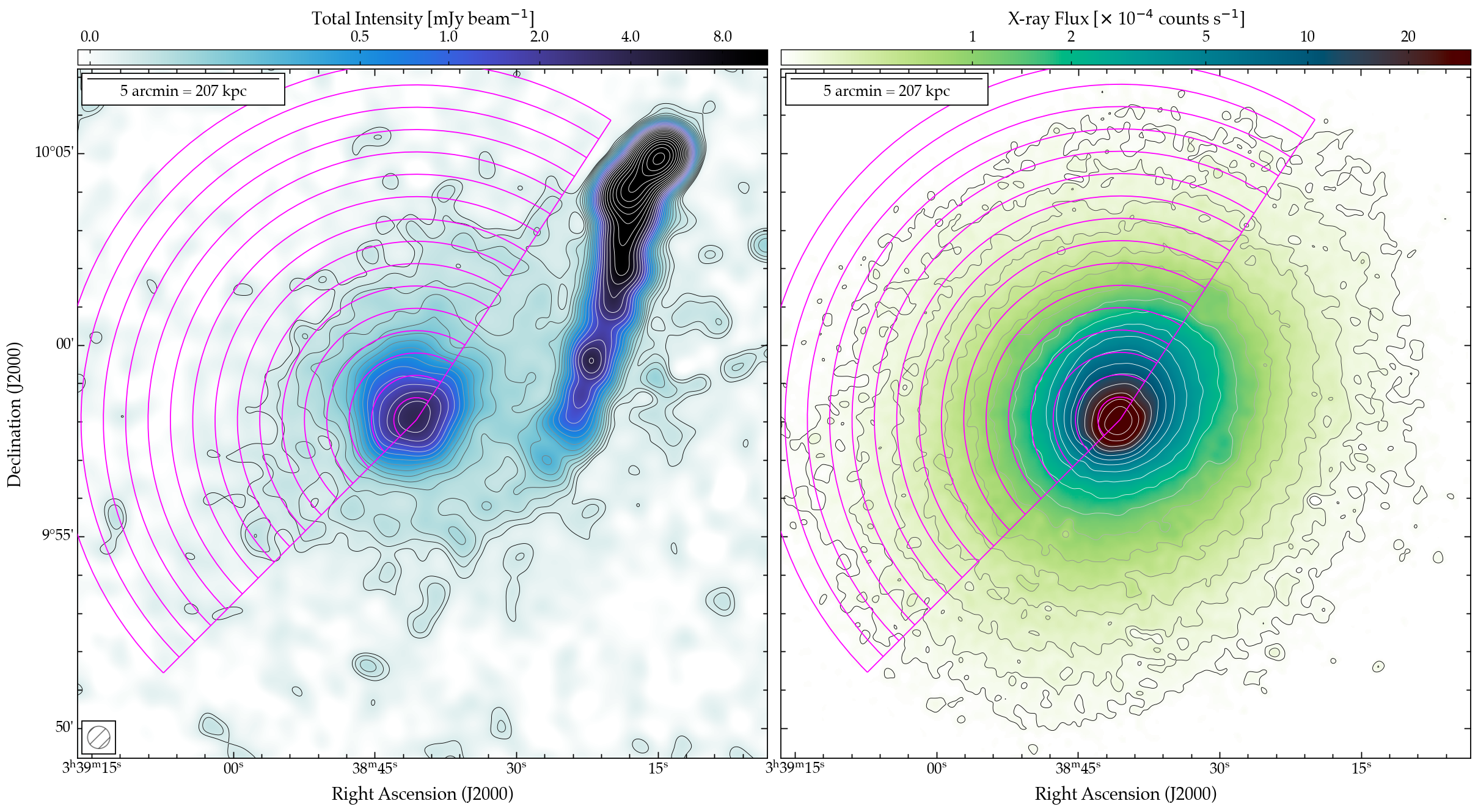}
    \caption{Regions used to derive the radial profiles in Fig.~\ref{fig:2A0335_radial_profile}. MeerKAT image (left) is as per Fig.~\ref{fig:2A0335_diffuse_35asec}; the \textit{XMM-Newton} surface brightness image (right) is as per Fig.~\ref{fig:2A0335_cluster_presub_xray}.}
    \label{fig:2A0335_wedges}
\end{figure*}

\begin{figure}
\sidecaption
    \includegraphics[width=0.45\linewidth]{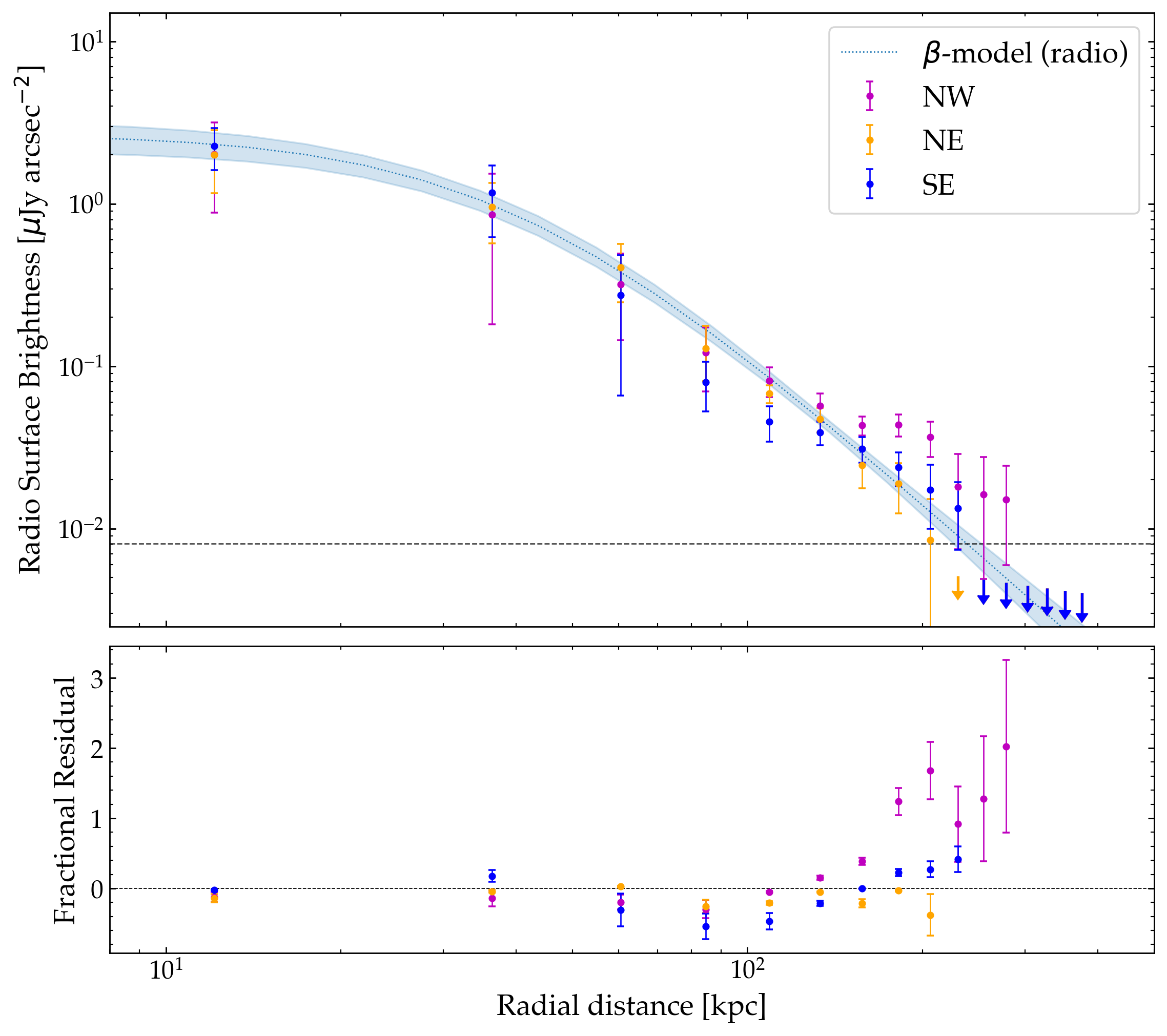}
    \caption{Radial profiles of the radio halo in 2A~0335+096 along different sectors (upper) as well as residuals compared to the best-fit global $\beta$-model (lower) from Fig.~\ref{fig:2A0335_radial_profile}.}
    \label{fig:2A0335_radial_profile_arcs}
\end{figure}

Here we show the regions used to quantify the thermal and non-thermal point-to-point correlations (Fig.~\ref{fig:2A0335_boxes}) and the radial bins used to explore the radial profiles (Fig.~\ref{fig:2A0335_wedges}). We also show the sector profiles in Fig.~\ref{fig:2A0335_radial_profile_arcs}.

\section{Monte-Carlo analysis of point-to-point correlations}\label{appendix_mcmc}

To investigate the robustness of our point-to-point $I_{\rm R}/I_{\rm X}$ correlation against an accidental ``sweet-spot'' choice of division between inner and outer regions, we performed a Monte-Carlo (MC) analysis. We defined a boundary circle based on a physically-motivated radius, using the apparent break in correlation between 1D X-ray temperature and $I_{\rm X}$ (or $I_{\rm R}$) shown in Fig.~\ref{fig:2A0335_PTP_T1D}.

\begin{figure*}
    \sidecaption
    \includegraphics[width=12cm]{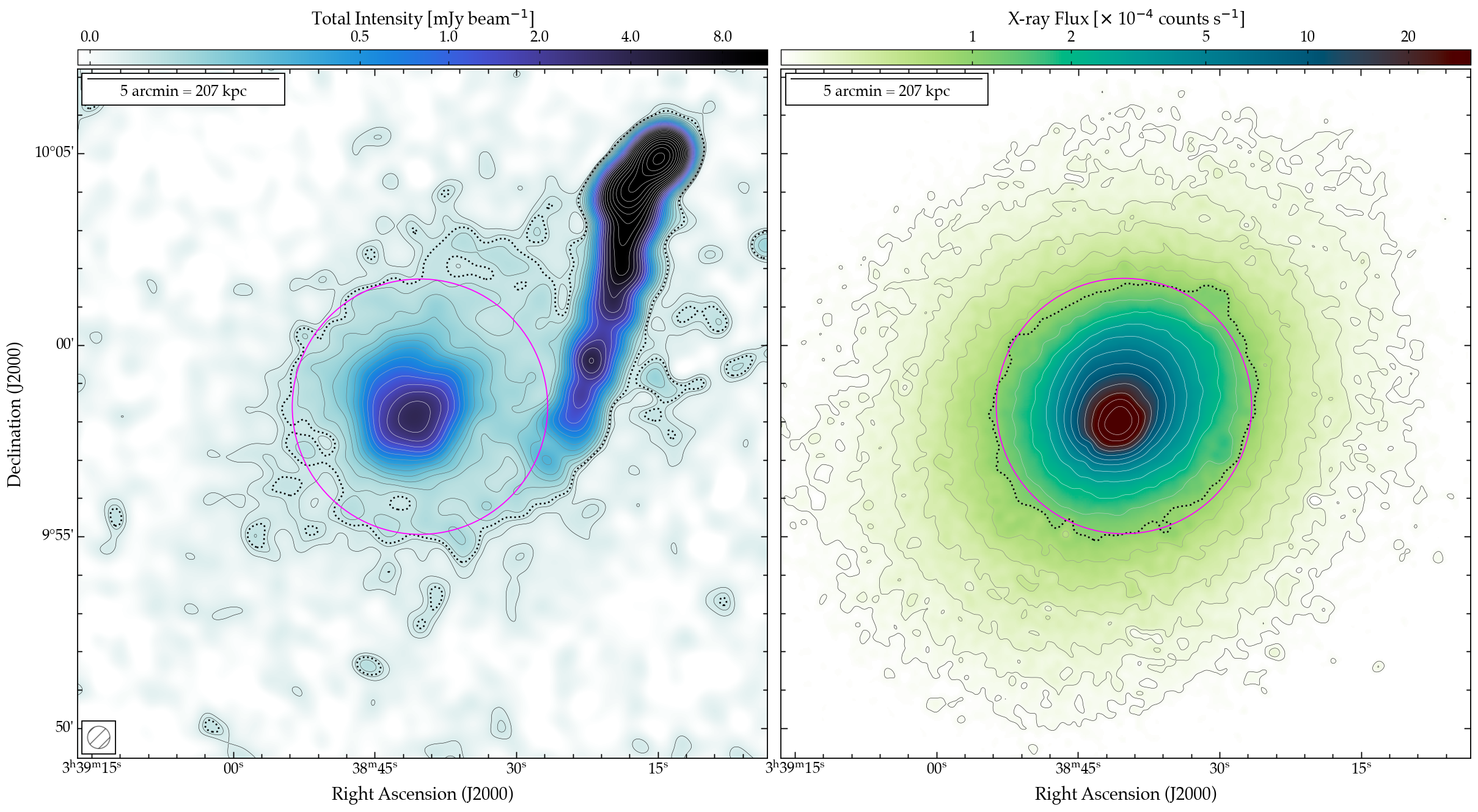}
    \caption{Initial boundary circle used for our MC analysis (magenta) and corresponding surface brightness contour used to delineate the inner and outer region sets (shown by the dotted contour). MeerKAT image (left) is as per Fig.~\ref{fig:2A0335_diffuse_35asec}; the \textit{XMM-Newton} surface brightness image (right) is as per Fig.~\ref{fig:2A0335_cluster_presub_xray}.}
    \label{fig:2A0335_mc_boundary}
\end{figure*}

From Fig.~\ref{fig:2A0335_PTP_T1D} we see an apparent transition from inner to outer trends at around $\log{I_{\rm X}} \simeq -7.2$ (or $\log{I_{\rm R}} \simeq -7.4$, although the break in $\log{I_{\rm X}}$ appears sharper). We defined our initial inner/outer boundary circle based on the threshold in $\log{I_{\rm X}}$, using a circular region that would most closely encompass this contour. This circular region is centred on $\rm (RA, \, Dec.)$ = $(54.6678, \, 9.9730)$ and has a radius of 200~arcsec. We show this circle, along with the corresponding surface brightness threshold, in Fig.~\ref{fig:2A0335_mc_boundary}; the same circle overlaid on the radio surface brightness map also closely encompasses the corresponding surface brightness threshold in $\log{I_{\rm R}}$, although due to the asymmetry of the diffuse emission the correspondence is less ``good''.

From this initial circle, we generated 100 realisations of boundary circles, perturbing the radius using a factor drawn from a Gaussian distribution with a central value of 1.0 and a standard deviation of 10\%. We then used the combined region set shown in Fig.~\ref{fig:2A0335_boxes} and created a new inner/outer region set per realisation, using the MC-generated boundary circles to define the inner/outer classification. For regions appearing on the boundary, we used a majority-pixel classification (i.e if the majority of pixels within a box were inside the circle, the whole box counts as inside the circle).

We then repeated our point-to-point correlation analyses on these realisations of inner/outer region sets. The results of this MC analysis are shown in Table~\ref{tab:correlation_results}, where we quote the 50th percentile of the fit results from the 100 realisations of region sets for all parameters.

\end{appendix}

\end{document}